\documentclass[aps,pre,reprint,longbibliography,showpacs,showkeys,amsmath,amssymb,twocolumn]{revtex4}
\usepackage{graphicx}
\usepackage{amsfonts}
\usepackage{url}
\usepackage{epstopdf}
\usepackage{color}
\usepackage{bm}
\usepackage[colorlinks=true,linkcolor=blue,citecolor=red]{hyperref}%

\usepackage{multirow}

\usepackage{bm}
\usepackage{amssymb}
\usepackage{amsmath}
\usepackage{ulem}
\usepackage{comment} 
\usepackage[dvipsnames]{xcolor}

\usepackage{color,soul}  
\setstcolor{red}         

\newcommand{\be}{\begin{equation}}
\newcommand{\ee}{\end{equation}}
\newcommand{\ba}{\begin{eqnarray}}
\newcommand{\ea}{\end{eqnarray}}
\newcommand{\besu}{\begin{subequations}}
\newcommand{\esu}{\end{subequations}}

\def\x{\bm{x}}

\DeclareMathOperator{\erfc}{erfc}

\def\x{\bm{x}}

\def\ve{\varepsilon}

\def\pa{\partial\Omega}

\def\I{{\mathbb I}}
\def\P{{\mathbb P}}
\def\R{{\mathbb R}}

\def\T{{\mathcal T}}

\def\D{{\mathcal D}}

\def\erfc{\mathrm{erfc}}

\begin{document}

\title{First-encounter time of two diffusing particles \\ in two- and three-dimensional confinement}

\author{F. Le Vot}
\affiliation{
Departamento de F\'{\i}sica and Instituto de Computaci\'on Cient\'{\i}fica Avanzada (ICCAEx) \\
Universidad de Extremadura, E-06071 Badajoz, Spain}

\author{S.~B. Yuste}
\affiliation{
Departamento de F\'{\i}sica and Instituto de Computaci\'on Cient\'{\i}fica Avanzada (ICCAEx) \\
Universidad de Extremadura, E-06071 Badajoz, Spain}

\author{E. Abad}
\affiliation{
Departamento de F\'{\i}sica Aplicada and Instituto de Computaci\'on Cient\'{\i}fica Avanzada (ICCAEx) \\
Centro Universitario de M\'erida \\ Universidad de Extremadura, E-06800 M\'erida, Spain}

\author{D.~S. Grebenkov}
\affiliation{
Laboratoire de Physique de la Mati\`{e}re Condens\'{e}e (UMR 7643), \\
CNRS -- Ecole Polytechnique, IP Paris, 91128 Palaiseau, France}

\begin{abstract}
The statistics of the first-encounter time of diffusing particles
changes drastically when they are placed under confinement.  In the
present work, we make use of Monte Carlo simulations to study the
behavior of a two-particle system in two- and three-dimensional
domains with reflecting boundaries.  Based on the outcome of the
simulations, we give a comprehensive overview of the behavior of the
survival probability $S(t)$ and the associated first-encounter time
probability density $H(t)$ over a broad time range spanning several
decades.  In addition, we provide numerical estimates and empirical
formulas for the mean first-encounter time $\langle \cal{T}\rangle $,
as well as for the decay time $T$ characterizing the monoexponential
long-time decay of the survival probability. Based on the distance
between the boundary and the center of mass of two particles, we
obtain an empirical lower bound $t_B$ for the time at which $S(t)$
starts to significantly deviate from its counterpart for the no
boundary case.  Surprisingly, for small-sized particles, the dominant
contribution to $T$ depends only on the total diffusivity $D=D_1+D_2$,
in sharp contrast to the one-dimensional case.  This contribution can
be related to the Wiener sausage generated by a fictitious Brownian
particle with diffusivity $D$.  In two dimensions, the first
subleading contribution to $T$ is found to depend weakly on the ratio
$D_1/D_2$.  We also investigate the slow-diffusion limit when $D_2 \ll
D_1$ and discuss the transition to the limit when one particle is a
fixed target.  Finally, we give some indications to anticipate when
$T$ can be expected to be a good approximation for $\langle
\cal{T}\rangle$.
\end{abstract}





\maketitle

\section{Introduction}

The first-encounter time (FET) of diffusing particles is one of the
central quantities characterizing diffusion-influenced reactions.
Smoluchowski first recognized the importance of the encounter step by
showing that the bimolecular reaction rate of two spherical particles
is proportional to their linear sizes and diffusivities
\cite{Smoluchowski1917}.  The original problem of two particles
diffusing in the three-dimensional Euclidean space is equivalent here
to the simpler problem of a single particle diffusing towards a static
target.  Smoluchowski solved the single-particle diffusion equation
and determined the survival probability and thus the probability
density of the first-passage time to the target, which is here
equivalent to the FET.  Since his seminal work, first-passage times to
static targets have been thoroughly investigated for various kinds of
diffusion processes, chemical kinetics, and geometric settings
\cite{Rice,Lauffenburger,Redner,Schuss,Metzler,Oshanin,Sano79,Agmon90,Levitz06,Condamin07,Condamin07b,Grebenkov07,Benichou10,Benichou10b,Grebenkov10a,Grebenkov10b,Benichou11,Bressloff13,Benichou14,Galanti16,Guerin16,Lanoiselee18,Grebenkov19,Grebenkov19d,Grebenkov20e,Grebenkov20a}.
In the case of a fixed small target embedded in an otherwise
reflecting boundary, one deals with the so-called narrow escape
problem, for which many asymptotic results have been derived
\cite{Holcman04,Schuss07,Benichou08,Pillay10,Cheviakov10,Cheviakov12,Rupprecht15,Grebenkov17,Grebenkov18b,Grebenkov19e}
(see also a review \cite{Holcman14}).  Numerous studies were also
dedicated to the problem of multiple particles diffusing on
translationally invariant (both finite and infinite) lattices or in
Euclidean spaces, which is relevant to chemical reactions involving
various species (see \cite{Kozak00a,Kozak00b,Nicolis01,Bentz03,Abad03,Abad05,Abad06,Moreau03}
and references therein).  In particular, the effect of inter-particle
interactions (e.g., excluded volume), and the cooperativity effect
when, for instance, several predators hunt for a prey, were analyzed
\cite{Szabo1988,Redner1999,Blythe2003,Bray04,Yuste08,Borrego09,Oshanin2009}.
Theoretical developments have been complemented by numerical
approaches, in which diffusion-reaction processes were modeled by
molecular dynamics or Monte Carlo simulations
\cite{McGuffee10,Ghost16,Samanta16}.  

In spite of this progress, the statistics of the FET between two
particles diffusing in confined domains remains poorly understood.  As
the translational symmetry is broken by the presence of a confining
boundary, the reduction of two diffusing particles to a single
particle diffusing towards a static target is prohibited.  One has
therefore to describe the dynamics of two particles inside a confining
domain, and the solution of diffusion-reaction equations becomes much
more sophisticated.  Amitai {\it et al.}  estimated the mean
first-encounter time (MFET) between two ends of a polymer chain by
computing the mean time for a Brownian particle to reach a narrow
domain in the polymer configuration space \cite{Amitai2012}.  Tzou
{\it et al.} studied the MFET for two particles diffusing on a
one-dimensional interval by solving the underlying diffusion equations
\cite{Tzou2014}.  In particular, they discussed the question whether a
mobile trap can improve capture times over a fixed trap.  Even for
such a simple geometric setting, an analytical solution of the problem
was not provided.  Agliari {\it et al.}  investigated the encounter
problem for random walks on branched structures, in particular, on
combs
\cite{Agliari2014,Agliari2016,Agliari2019}.  More recently, Lawley and
Miles computed the MFET for a very general diffusion model with many
{\it small} targets that can diffuse either inside a three-dimensional
domain, or on its two-dimensional boundary, their diffusivities can
stochastically fluctuate, while their reactivity can be stochastically
gated \cite{Lawley19}.  Nayak {\it et al.}  investigated the capture
of a diffusive prey by multiple predators in confined space via
intenstive Monte Carlo simulations \cite{Nayak20}.  In particular,
they focused on the characteristic timescale associated with rare
capture events and its dependence on the number of searchers, the
relative diffusivity of the target with respect to the searcher, and
the system size.  In our former paper
\cite{PartI}, we brought some analytic insights into the influence of
confinement onto the distribution of the FET in one-dimensional
settings, namely, for two particles diffusing on the half-line or on
an interval.  As discussed below, the problem of two particles could
be mapped here onto an equivalent problem of a single particle
diffusing on a planar region (a wedge or a rectangle) and then solved
exactly.

In this companion paper, we extend our analysis to two- and
three-dimensional confining domains.  We consider two Brownian
particles $A$ and $B$ diffusing inside a bounded domain with
reflecting boundary, until their encounter that triggers an
instantaneous chemical reaction: $A + B \rightarrow C$.  We
investigate the survival probability, i.e., the probability of both
particles not having met up to a given time $t$.  The survival
probability can be interpreted as the fraction of particles still
reactive at time $t$ with respect to the initial number of particles,
and it determines other important quantities such as the probability
density of the FET (whence its mean value and higher order moments
follow, as well as the reaction rate).

The paper is organized as follows.  In Sec.~\ref{sec:equations}, we
formulate the diffusion-reaction problem and summarize the main known
theoretical results that are relevant for our study.  The Monte Carlo
simulations and the statistical tools for analysis of the survival
probability and the FET probability density for two particles inside a
disk and a sphere with reflecting boundary are described in
Sec. \ref{sec:MC_general}.  The analysis in two dimensions is
developed for the particular case of a single particle in the search
for a fixed target (Sec.~\ref{Sec:Fixed_Target}), for two identical
diffusing particles (Sec.~\ref{Sec:Identical_Particles}), and for two
particles with different diffusivities (Sec.~\ref{Sec:Transition}).
Extensions to the three-dimensional case are presented in
Sec.~\ref{Sec:3d}, while the main conclusions are summarized in
Sec.~\ref{Sec:Conclusions}.  The appendix describes the details of
Monte Carlo simulations.

\section{Summary of some known theoretical results}
\label{sec:equations}

In this section, we summarize some theoretical results on the
first-encounter time in two- and three-dimensional space.  Even though
these results are known, they are dispersed in the literature and not
easily accessible.  A summary of results for one-dimensional settings
was provided in \cite{PartI}.

\subsection{Two diffusing particles}

We consider two spherical particles of radii $\rho_1$ and $\rho_2$,
started from prescribed points $\x_1$ and $\x_2$ and diffusing with
diffusion coefficients $D_1$ and $D_2$ in a $d$-dimensional Euclidean
domain $\Omega \subset \R^d$ with a smooth reflecting boundary $\pa$.
The first-encounter time $\T$ of these particles is a random variable
characterized by the cumulative probability distribution, $\P\{ \T <
t\}$, or, equivalently, by the survival probability $S(t|\x_1,\x_2) =
\P\{ \T > t\}$.  As the encounter depends on positions of both particles,
it is natural to consider their joint dynamics in the phase space
$\Omega \times \Omega$, which is governed by the second-order
differential operator
\begin{equation}
\D = - \bigl(D_1 \Delta_{\x_1} + D_2 \Delta_{\x_2} \bigr),
\end{equation}
where $\Delta_{\x_i}$ is the Laplace operator acting on $\x_i$.  The
survival probability satisfies the joint diffusion equation:
\begin{equation}  \label{eq:Sgeneral_def}
\frac{\partial S}{\partial t} = - \D \, S  \qquad (\x_1,\x_2)\in \Omega \times \Omega,
\end{equation}
subject to the initial condition $S(t=0|\x_1,\x_2) = 1$.  As the
boundary $\pa$ of the confining domain $\Omega$ is reflecting (there
is no net diffusive flux across the boundary), the Neumann boundary
condition applies for both particles:
\begin{subequations}
\begin{align}
 \frac{\partial S}{\partial n_1}  & = 0   \qquad (\x_1,\x_2) \in \pa \times \Omega,\\
 \frac{\partial S}{\partial n_2}  & = 0   \qquad (\x_1,\x_2) \in \Omega \times \pa,
\end{align}
\end{subequations}
where $\partial/\partial n_i$ is the normal derivative at the boundary
point $\x_i$ oriented outward $\Omega$.  As we are interested in the
first encounter, the Dirichlet boundary condition is imposed whenever
the particles are at contact, i.e., within the distance $|\x_1 - \x_2|
= \rho$:
\begin{equation}  \label{eq:SinGamma}
S = 0  \qquad (\x_1,\x_2) \in \Gamma ,
\end{equation}
where $\Gamma = \{(\x_1,\x_2)\in \Omega\times \Omega ~:~ |\x_1 - \x_2|
= \rho\}$, with
\begin{equation}
\rho = \rho_1 + \rho_2 .
\end{equation}
In other words, the first-encounter time of two diffusing particle is
equivalent to the first-passage time of a single diffusive process
$(X^{(1)}_t,X^{(2)}_t)$, describing the motion of these particles, to
the target $\Gamma$.
The survival probability determines the probability density of the
FET,
\begin{equation}  \label{eq:H_def}
H(t|\x_1,\x_2) = - \frac{\partial S(t|\x_1,\x_2)}{\partial t} \,,
\end{equation}
as well as the moments (if they exist):
\begin{equation}  \label{eq:moments_def}
\langle \T^k \rangle = \int\limits_0^{\infty} dt \, t^k \, H(t|\x_1,\x_2)
= k \int\limits_0^{\infty} dt \, t^{k-1}\, S(t|\x_1,\x_2),  
\end{equation}
with $k = 1,2,\ldots$  In particular, the MFET is the area below the
survival probability curve:
\begin{equation}  \label{eq:MFPT}
\langle \T \rangle = \int\limits_0^{\infty} dt \, S(t|\x_1,\x_2).
\end{equation}
From Eqs.~\eqref{eq:Sgeneral_def}-\eqref{eq:SinGamma} and
Eq. \eqref{eq:moments_def}, one also finds that the moments $ \langle
\T^k \rangle$ (if they exist) satisfy the well-known hierarchy of PDEs
\begin{equation}  \label{eq:TavgEq}
\D \langle \T^k \rangle = k \langle \T^{k-1} \rangle   \qquad (\x_1,\x_2)\in \Omega \times \Omega,
\end{equation}
with
\begin{subequations}
\begin{align}
 \frac{\partial \langle \T^k \rangle}{\partial n_1}  & = 0   \qquad (\x_1,\x_2) \in \pa \times \Omega,\\
 \frac{\partial \langle \T^k \rangle}{\partial n_2}  & = 0   \qquad (\x_1,\x_2) \in \Omega \times \pa,
\end{align}
\end{subequations}
and
\begin{equation}  \label{eq:TavginGamma}
\langle \T^k \rangle = 0  \qquad (\x_1,\x_2) \in \Gamma .
\end{equation}

For any {\it bounded} domain $\Omega$, the solution of the boundary
value problem \eqref{eq:Sgeneral_def} -- \eqref{eq:SinGamma} can be
formally expanded over the eigenfunctions of the governing diffusion
operator $\D$ in Eq. (\ref{eq:Sgeneral_def}):
\begin{equation}   \label{eq:S_spectral}
S(t|\x_1,\x_2) = \sum\limits_{n=1}^\infty e^{-\Lambda_n t} \, U_n(\x_1,\x_2) \hspace*{-2mm} 
\int\limits_{\Omega\times\Omega} \hspace*{-2mm}   d\x'_1 \, d\x'_2 \, U_n^*(\x'_1,\x'_2),
\end{equation}
where the asterisk denotes the complex conjugate, $\Lambda_n$ are the
eigenvalues and $U_n(\x_1,\x_2)$ are the
$L_2(\Omega\times\Omega)$-normalized eigenfunctions of $\D$: $\D U_n =
\Lambda_n U_n$ ($n = 1,2,\ldots$) \cite{Gardiner}.  The eigenvalues
are positive, have units of inverse time, and can be enumerated in the
ascending order: $0 \leq \Lambda_1 \leq \Lambda_2 \leq \ldots \nearrow
+\infty$, whereas the eigenfunctions form a complete basis allowing
for such spectral expansions.  In particular, the survival probability
and the FET density exhibit an exponential decay at long times,
\begin{equation}  \label{eq:S_expon}
S(t|\x_1,\x_2) \propto e^{-t/T}  \qquad (t\to\infty),
\end{equation}
with the decay time 
\begin{equation}  \label{eq:T_Lambda1} 
T = \frac{1}{\Lambda_1} \,,
\end{equation}
determined by the smallest eigenvalue $\Lambda_1$.  We emphasize that
$T$ does not depend on the starting points $\x_1$ and $\x_2$.  The
exponential decay implies that all positive moments of $\T$ are
finite.

In the previous paper \cite{PartI}, we discussed how this general
description can be applied in one-dimensional settings, in which
$\Omega\times\Omega$ is a planar region and $\Gamma$ is either a
half-line or an interval.  In higher dimensions ($d \geq 2$), $\Gamma$
is a $(2d-1)$-dimensional region (of nontrivial shape) in a
$2d$-dimensional domain $\Omega\times\Omega$ that makes analytical
solutions generally unfeasible.  An exception is the case of diffusion
in free space, $\Omega = \R^d$, for which the change of coordinates
simplifies the problem and allows one to get the solution:

(i) In three dimensions, the solution was found by Smoluchowski
\cite{Smoluchowski1917},
\begin{equation}  \label{eq:S_free3D}
S_{\rm free}(t|\x_1,\x_2) = 1 - \frac{\rho}{r} \erfc \left( \frac{r-\rho}{\sqrt{4 D t}} \right),
\end{equation}
where
\begin{equation*}
r = |\x_1 - \x_2|
\end{equation*}
is the initial distance between the centers of two particles,
\begin{equation}
D = D_1 + D_2,
\end{equation}
and $\erfc(z)$ is the complementary error function.  The probability
density of the FET is
\begin{equation}  \label{eq:H_free3D}
H_{\rm free}(t|\x_1,\x_2) = \frac{\rho}{r} \, \frac{r-\rho}{\sqrt{4\pi D t^3}} \exp \left(- \frac{(r-\rho)^2}{4 D t} \right).
\end{equation}

(ii) In two dimensions, there is an explicit formula for the Laplace
transform of the survival probability:
\begin{align}  \nonumber
\tilde{S}_{\rm free}(p|\x_1,\x_2) & = \int\limits_0^\infty dt \, e^{-pt} \, S_{\rm free}(t|r)  \\   \label{SFree_Target_Laplace}
& = \frac{1}{p} \biggl( 1 - \frac{ K_0 ( r \sqrt{p/D})}{ K_0 (\rho \sqrt{p/D})} \biggr),
\end{align}
where $K_{\nu}(\cdot)$ is the $\nu$th-order modified Bessel function
of the second kind.  The inverse Laplace transform can be expressed as
\cite{Grebenkov18b}
\begin{align}  \label{eq:Sfree_2d}
S_{\rm free}(t|\x_1,\x_2) & = \frac{2}{\pi} \int\limits_0^\infty \frac{dq}{q} \, e^{-Dtq^2}  \\  \nonumber
& \times \frac{Y_0(qr) J_0(q\rho)
- J_0(qr) Y_0(q\rho)}{J_0^2(q\rho) + Y_0^2(q\rho)} \,,
\end{align}
where $J_{\nu}(\cdot)$ and $Y_{\nu}(\cdot)$ are respectively the
$\nu$th-order Bessel functions of the first and second kind.  One also
gets
\begin{align}  \label{eq:Hfree_2d}
H_{\rm free}(t|\x_1,\x_2) & = \frac{2D}{\pi} \int\limits_0^\infty dq \, q \, e^{-Dtq^2}  \\  \nonumber
& \times \frac{Y_0(qr) J_0(q\rho)
- J_0(qr) Y_0(q\rho)}{J_0^2(q\rho) + Y_0^2(q\rho)} \,.
\end{align}
This integral representation allows for a rapid numerical computation
of $H_{\rm free}(t|\x_1,\x_2)$.  Levitz {\it et al.}  proposed an
explicit approximation for this density \cite{Levitz2008}, but it is
only valid when $r$ is close to $\rho$ (see the discussion in the
Supplemental Information of \cite{Grebenkov18b}).  This density
exhibits an extremely slow decay at long times:
\begin{equation}  \label{eq:Hfree_2d_decay}
H_{\rm free}(t|\x_1,\x_2) \simeq \frac{2(r/\rho - 1)}{t \ln^2(2Dt/\rho^2)}  \qquad (t\to\infty),
\end{equation}
as well as the survival probability:
\begin{equation}  \label{eq:Sfree_2d_decay}
S_{\rm free}(t|\x_1,\x_2) \simeq \frac{2(r/\rho - 1)}{\ln(2Dt/\rho^2)}  \qquad (t\to\infty).
\end{equation}

\subsection{Single particle diffusing towards a static target}

Due to mathematical challenges encountered in the analysis of the
above problem \eqref{eq:Sgeneral_def}-\eqref{eq:SinGamma} for two
diffusing particles in a confinement, most former theoretical works
dealt with a much simpler setting, in which one particle diffuses
towards an immobile particle considered as a static target or a sink
\cite{Rice,Lauffenburger,Redner,Schuss,Metzler,Oshanin,Sano79,Agmon90,Levitz06,Condamin07,Condamin07b,Grebenkov07,Benichou10,Benichou10b,Grebenkov10a,Grebenkov10b,Benichou11,Bressloff13,Benichou14,Galanti16,Guerin16,Lanoiselee18,Grebenkov19,Grebenkov19d,Grebenkov20e}.
This problem is equivalent to diffusion of a single point-like
particle with diffusivity $D_1 = D$ inside a modified domain
$\Omega'$:
\begin{equation}
\Omega' = \{ \x_1 \in \Omega ~:~ |\x_1 - \pa| > \rho_1,  ~ |\x_1 - \x_2| > \rho\} ,
\end{equation}
where $\x_2$ is the fixed position of the target (i.e., the second
particle with diffusivity $D_2 = 0$), and $|\x_1 - \pa|$ is the
Euclidean distance from $\x_1$ to the boundary $\pa$.  In other words,
the diffusing particle of radius $\rho_1$ cannot get closer to the
boundary $\pa$ of the confining domain $\Omega$ than by a distance
$\rho_1$, and cannot overlap with the fixed target of radius $\rho_2$.
The survival probability satisfies the ordinary diffusion equation,
\begin{equation}  \label{eq:S1_diff}
\frac{\partial S}{\partial t} = D \Delta_{\x_1} S  \qquad \x_1 \in \Omega' ,
\end{equation}
subject to the initial condition $S(t = 0|\x_1) = 1$ and the mixed
boundary conditions:
\begin{align}  \label{eq:S1_BC}
\frac{\partial S}{\partial n} & = 0  \qquad \x_1 \in \pa' , \\   \label{eq:S1_BC2}
S & = 0  \qquad \x_1 \in \Gamma' ,
\end{align}
where $\pa' = \{ \x_1 \in \Omega ~:~ |\x_1 - \pa| = \rho_1\}$ is the
reflecting boundary of the shrunk confining domain $\Omega'$, and
$\Gamma' = \{ \x_1 \in \Omega ~:~ |\x_1 - \x_2| = \rho\}$ is the
encounter region (for the sake of simplicity, we assumed that $|\x_2 -
\pa| > \rho$, i.e. $\pa'$ and $\Gamma'$ are disjoint; but more general
settings can be considered as well.

As the boundary value problem (\ref{eq:S1_diff} -- \ref{eq:S1_BC2})
has been thoroughly investigated and reviewed in the past, we only
summarize several results that will be relevant for our analysis.  For
any bounded domain $\Omega'$, the spectrum of the Laplace operator is
discrete, and the solution of (\ref{eq:S1_diff}, \ref{eq:S1_BC})
admits a general spectral expansion \cite{Redner,Gardiner}
\begin{equation}    \label{eq:St_general}
S(t|\x_1,\x_2) = \sum_{n=1}^{\infty} u_n(\x_1;\x_2) \, e^{ - t \lambda_n(\x_2)} \int_{\Omega'} d\x' ~ u_n^* (\x';\x_2),
\end{equation}
where $\lambda_n$ and $u_n$ are the $n$th eigenvalue and
$L_2(\Omega')$-normalized eigenfunction of the diffusion operator $\D'
= - D \Delta_{\x_1}$, both depending on the position $\x_2$ of the
static target through the shape of $\Omega'$.  To avoid confusion, we
distinguish the eigenpairs $(\lambda_n,u_n)$ from $(\Lambda_n,U_n)$
used in the case of two diffusing particles.  The eigenvalues can be
ordered such as $0 < \lambda_1 \leq \lambda_2 \leq \ldots \nearrow
+\infty$.  In particular, the survival probability decays
exponentially at long times,
\begin{equation}  \label{eq:St_decay}
S(t|\x_1,\x_2) \propto e^{ - t/T(\x_2)}  \qquad (t\to \infty)\,,
\end{equation}
with the decay time $T(\x_2)$ determined by the smallest eigenvalue:
\begin{equation}  \label{eq:Tx2}
T(\x_2) = \frac{1}{\lambda_1(\x_2)}  \,,
\end{equation}
where we highlighted the dependence on the target position $\x_2$, in
contrast to the case (\ref{eq:T_Lambda1}) of two diffusing particles.

\subsubsection*{Concentric domains}

The eigenvalues and eigenfunctions of the diffusion operator are in
general not known explicitly.  One of few exceptions is the case when
$\Gamma$ and $\pa$ are concentric circles or spheres of radii $\rho$
and $R$, respectively (i.e., $\x_2 = 0$).  In this case, the
rotational symmetry of $\Omega'$ implies that $S(t|\x_1,\x_2)$ depends
on $\x_1$ only via its radial coordinate, $r = |\x_1 - \x_2| =
|\x_1|$, that allows one to solve Eq. (\ref{eq:S1_diff}) in the
Laplace space \cite{Redner} (see also
\cite{Levitz2008,Grebenkov2017,Grebenkov2018}).  Denoting by
$\tilde{S}$ and $\tilde{H}$ the Laplace transforms of $S$ and $H$,
respectively, the solution can be written as
\begin{equation}  \label{eq:tildeS_23d}
\tilde{S}(p|\x_1,\x_2) = \frac{1}{p} \left[ 1 - \tilde{H}(p|\x_1,\x_2) \right],
\end{equation}
with
\begin{equation} \label{eq:tildeH_d}
\tilde{H}(p|\x_1,\x_2) = (\rho/r)^{\nu} 
\frac{I_{\nu+1}(z \bar{R}) K_\nu (z r) + K_{\nu+1}(z \bar{R}) I_\nu (z r)}{I_{\nu+1}(z \bar{R}) K_\nu (z \rho) + K_{\nu+1}(z \bar{R}) I_\nu (z \rho)} \,,
\end{equation}
where $\nu = d/2-1$, $z=\sqrt{p/D}$, $I_{\nu}(\cdot)$ is the
$\nu$th-order modified Bessel function of the first kind, and
\begin{equation*}
\bar{R} = R - \rho_1 .
\end{equation*}
More explicitly, one has
\begin{equation} \label{eq:tildeH_2d}
\tilde{H}(p|\x_1,\x_2) = \frac{I_1(z \bar{R}) K_0 (z r) + K_1(z \bar{R}) I_0 (z r)}{I_1(z \bar{R}) K_0 (z \rho) + K_1(z \bar{R}) I_0 (z \rho)} 
\end{equation}
in two dimensions, and
\begin{equation}  \label{eq:tildeH_3d} 
\tilde{H}(p|\x_1,\x_2) = \frac{\rho}{r} \,
\frac{\bar{R} z \cosh(\bar{R}-r)z - \sinh(\bar{R}-r)z}{\bar{R} z \cosh (\bar{R}-\rho)z - \sinh (\bar{R}-\rho)z } 
\end{equation}
in three dimensions.

The inverse Laplace transform of $\tilde{H}(p|\x_1,\x_2)$ can be
performed by means of the residue theorem.  These expressions
determine all the moments of the FET, in particular,
\begin{align}  \label{eq:MFPT_concentric}
\langle \T \rangle & = \frac{\bar{R}^2 \ln(r/\rho)}{2D} - \frac{r^2 - \rho^2}{4D} \qquad (d = 2), \\
\label{3d_MFPT_Target}
\langle \T \rangle & = \frac{\bar{R}^3 (r - \rho)}{3 D r \rho} - \frac{r^2 - \rho^2}{6 D} \qquad (d = 3).
\end{align}
The eigenvalues $\lambda_n$ contributing to the survival probability
and to the FET probability density are related to the poles of
Eqs. (\ref{eq:tildeH_2d}, \ref{eq:tildeH_3d}):
\begin{equation}
\lambda_n = \alpha_n^2/\bar{R}^2 ,
\end{equation}
where $\alpha_n$ are positive solutions of
\begin{equation} \label{eq:alpha_n}
J_1(\alpha_n) Y_0(\alpha_n \rho/\bar{R}) - Y_1(\alpha_n) J_0(\alpha_n \rho/\bar{R}) = 0
\end{equation}
in two dimensions, and of
\begin{equation}  \label{eq:alpha_3d}
\tan [\alpha_n(1-\rho/\bar{R})] = \alpha_n
\end{equation}
in three dimensions.  In the small target limit, $\rho \to 0$, the
smallest eigenvalue $\lambda_1$ vanishes as:
\begin{equation}
\lambda_1 \simeq \frac{D}{\bar{R}^2} \left\{ \begin{array}{l l}  2/\ln(\bar{R}/\rho) & (d=2), \\  3\rho/\bar{R} & (d=3), \\  \end{array} \right.
\end{equation}
so that the decay time $T$ from Eq. (\ref{eq:Tx2}) increases as
\begin{equation}  \label{eq:Tsmall}
T \simeq \frac{\bar{R}^2}{D d} \left\{ \begin{array}{l l}  \ln(\bar{R}/\rho) & (d=2), \\  \bar{R}/\rho & (d=3) \\  \end{array} \right.
\end{equation}
in the leading order.  It is instructive to compare the time $T$ with
the mean first-passage time (MFPT) $\langle \T \rangle$ given by
Eqs. \eqref{eq:MFPT_concentric}, \eqref{3d_MFPT_Target}:
\begin{equation}
\frac{2dD}{\bar{R}^2} \bigl(T - \langle \T\rangle\bigr) \simeq 
\left\{ \begin{array}{l l}  2\ln(\bar{R}/r) + (r/\bar{R})^2 & (d=2), \\  
2(\bar{R}/r) + (r/\bar{R})^2 & (d=3). \\  \end{array} \right.
\end{equation}
One sees that the decay time $T$ always exceeds $\langle \T\rangle$,
and that the difference between these two quantities is minimal at $r
= \bar{R}$.  This is a signature of the prevalence of long
trajectories in the behavior of the long time decay; note that the
MFET may be smaller {\it or} larger than the decay time in the case of
two diffusing particles (see Sec. \ref{Sec:Transition}).

In the limit $R\to \infty$ of an infinite domain $\Omega$,
Eqs. (\ref{eq:tildeS_23d}) and (\ref{eq:tildeH_2d}) lead to
Eq. (\ref{SFree_Target_Laplace}), whereas the inverse Laplace
transform of the limit of Eqs. (\ref{eq:tildeS_23d}) and
(\ref{eq:tildeH_3d}) yields Eq. (\ref{eq:S_free3D}).

\subsubsection*{Small-target limit}

For a {\it small} fixed target in an arbitrary bounded domain
$\Omega'$, the asymptotic behavior of the smallest eigenvalue of the
Laplace operator has been thoroughly investigated (see
\cite{Kolokolnikov05,Cheviakov11} and references therein).

For a confining disk of radius $\bar R$, one has \cite{Kolokolnikov05}
\begin{equation}
\lambda_1 = \frac{2\pi \nu D}{|\Omega'|} - \frac{4\pi^2 \nu^2}{|\Omega'|} G(\x_2,\x_2) + O(\nu^3),
\end{equation}
where $\nu = -1/\ln \ve$, $\ve=\rho/\bar{R}$ is the dimensionless size
of the target, $|\Omega'|$ is the area of the shrunk domain $\Omega'$,
$\x_2$ is the location of the target, and $G(\x_2,\x_2)$ is the
regular part of the Neumann Green's function:
\begin{equation}
G(\x_2,\x_2) = - \frac{1}{2\pi} F_2(|\x_2|/\bar{R}) ,
\end{equation}
where
\begin{equation}
F_2(z) = \frac34 + \ln(1-z^2) - z^2 ,
\end{equation}
so that
\begin{equation}  \label{eq:lambda0_disk}
\lambda_1 \simeq \frac{2\nu D}{\bar{R}^2} \biggl(1 + \nu F_2(|\x_2|/\bar{R})\biggr) .
\end{equation}
The decay time is then
\begin{equation}  \label{eq:Tsmall_2d}
T(\x_2) \simeq \frac{\bar{R}^2 \ln(\bar{R}/\rho)}{2D} \biggl(1 + \frac{F_2(|\x_2|/\bar{R})}{\ln(\bar{R}/\rho)} \biggr)^{-1} \quad (\rho\ll \bar{R}) \,.
\end{equation}
This expression refines Eq. (\ref{eq:Tsmall}), which corresponds to
$|\x_2| = 0$, with $F_2(0) = 3/4$.  In turn, the above asymptotic
relation is not applicable when $|\x_2|$ approaches $\bar{R}$ (i.e.,
when the target is too close to the boundary) because of the
logarithmic divergence of the correction term (see below the
asymptotic form of the MFPT, which remains well defined in this
limit).

In three dimensions, one has \cite{Cheviakov11}
\begin{equation}
\lambda_1 = D\bigl(\ve \lambda^{(1)} + \ve^2 \lambda^{(2)} + O(\ve^3)\bigr) ,
\end{equation}
where 
\begin{equation}
\lambda^{(1)} = \frac{4\pi C}{|\Omega'|} = \frac{3}{\bar{R}^3} \,,
\end{equation}
and $C$ is the capacitance of the target of unit size (which is equal
to $1$ in the case of an spherical target).  The next-order correction
$\lambda^{(2)}$ is again expressed in terms of the regular part of the
Neumann Green function.  For a spherical confining domain of radius
$\bar{R}$, one has
\begin{align}  \nonumber
\lambda_1 & = \frac{4\pi \rho D}{|\Omega'|} - \frac{16\pi^2 \rho^2}{|\Omega'|} G(\x_2,\x_2) + O(\rho^3) \\
& = \frac{3\rho D}{\bar{R}^3} \biggl(1 - \frac{\rho}{\bar{R}} F_3(|\x_2|/\bar{R}) + O(\rho^2)\biggr) ,
\end{align}
where
\begin{equation}
F_3(z) = \frac{1}{1-z^2} - \ln (1-z^2) + z^2 - \frac{14}{5} \,.
\end{equation}
As a consequence, the decay time behaves as
\begin{equation}  \label{eq:Tsmall_3d}
T(\x_2) \simeq  \frac{\bar{R}^3}{3D\rho} \biggl(1 - \frac{\rho}{\bar{R}} F_3(|\x_2|/\bar{R})\biggr)^{-1}  \quad (\rho \ll \bar{R})\,.
\end{equation}
This expression refines Eq. (\ref{eq:Tsmall}), which corresponds to
$|\x_2| = 0$, with $F_3(0) = -9/5$, and thus
\begin{equation}
\lambda_1 = \frac{3\rho D}{\bar{R}^3} \biggl(1 + \frac{9}{5} \, \frac{\rho}{\bar{R}} + O(\rho^2)\biggr) .
\end{equation}
Note that this result agrees with the direct asymptotic analysis of
the smallest eigenvalue obtained as $\lambda_1 = D
\alpha_1^2/\bar{R}^2$, where $\alpha_1$ is the smallest strictly
positive solution of Eq. (\ref{eq:alpha_3d}), see
\cite{Grebenkov2018}.  Again, the opposite limit $|\x_2|\to \bar{R}$
yields the divergent correction term, and thus is not applicable.

The asymptotic behavior of the MFPT to a small target located on the
boundary of the domain was given in Ref. \cite{Benichou08}:
\begin{equation}
\langle \T\rangle \simeq \frac{|\Omega'|}{2\pi D} \times \left\{ \begin{array}{l l} \ln(r/\rho) & (d=2), \\
\frac{\Gamma(d/2)}{\pi^{d/2-1}} \bigl(\rho^{2-d} - r^{2-d}\bigr) & (d\geq 3) , \\  \end{array} \right.
\end{equation}
where $r = |\x_1 - \x_2|$ is the distance between the target and the
starting position of the diffusing particle.  For instance, one gets
\begin{equation}
\langle \T\rangle \simeq \frac{\bar{R}^2 \ln(\bar{R}/\rho)}{2D} \biggl(1 + \frac{\ln(r/\bar{R})}{\ln(\bar{R}/\rho)} \biggr)
\end{equation}
for a disk of radius $\bar{R}$, and
\begin{equation}
\langle \T\rangle \simeq \frac{\bar{R}^3}{3D\rho} \biggl(1 - \frac{\rho}{r} \biggr)
\end{equation}
for a sphere of radius $\bar{R}$.  While the leading terms in both
expressions are identical with those in Eqs. (\ref{eq:Tsmall_2d},
\ref{eq:Tsmall_3d}), the MFPTs depend on the positions of both
particles (the searcher and the target), whereas the decay time $T$
depends only on the position of the target.

\section{Monte Carlo simulations}
\label{sec:MC_general}

In this work, we undertake a systematic study of the FET distribution
in two and three-dimensional domains.  We restrict our analysis to two
particles of identical radii:
\begin{equation}
\rho_1 = \rho_2 = \rho/2 .
\end{equation}
We fix length units by setting $\rho = 1$.  In turn, we vary other
parameters such as diffusion coefficients ($D_1$, $D_2$), the initial
positions of particles ($\x_1$, $\x_2$), and the size of the
confinement ($R$).  While both the mathematical analysis of the
boundary value problem \eqref{eq:Sgeneral_def} -- \eqref{eq:SinGamma}
and the associated numerical simulations can be performed for
particles of arbitrary size (under the evident geometric constraint
$2\rho < R$), we restrict our study to the case of relatively small
particles: $\rho \ll R$.  Even though the limit of strong confinement
(particle diameter comparable to domain diameter) is also interesting
for applications, we will focus on systems with relatively small
particles.

For a given set of parameters, we simulated individual trajectories of
two diffusing particles in confinement and computed their FET $\T_i$
in each run $i$ (see Appendix for technical details).  To avoid
exceedingly long trajectories, we introduced a cut-off time $t_{\rm
cut}$, at which the simulation was stopped, even if two particles had
not met.  The cut-off time was large enough to ensure that $S(t_{\rm
cut}|\x_1,\x_2)$ was very small so that the cut-off did not influence
the results (see below).  The simulation was repeated $N = 10^6$ times
to get a good enough FET statistics and to access the long-time
behavior of the survival probability.  The empirical curves of
$S(t|\x_1,\x_2)$ were obtained by dividing the number of realizations
with $\T_i>t$ by $N$, whereas the empirical curves of $H(t|\x_1,\x_2)$
were obtained as renormalized histograms obtained from the values of
$\T_i$.

Even though we will generally display $S(t|\x_1,\x_2)$ and
$H(t|\x_1,\x_2)$ for a broad range of timescales, the data
corresponding to large times exhibit high statistical uncertainties.
In fact, since we use $N=10^6$ realizations, values of, say,
$S(t)\lesssim 10^{-4}$, were estimated with a relatively small number
of outcomes and have thus to be taken with care.  There exist
efficient methods for improving the statistical accuracy of rare
events in Monte Carlo simulations.  For instance, Nayak {\it et al.}
implemented one such method to access the long-time behavior of the
survival probability \cite{Nayak20}.  As our focus is on the study of
the whole distribution of the FET, we keep using the basic Monte Carlo
scheme.

The simulation results are systematically compared to the available
analytical results and approximations.  The survival probability and
the FET probability density in the no-boundary case, $S_{\rm
free}(t|\x_1,\x_2)$ and $H_{\rm free}(t|\x_1,\x_2)$, are given by
explicit formulas (\ref{eq:S_free3D}, \ref{eq:H_free3D}) in the
three-dimensional case; in turn, a numerical integration of
Eqs. (\ref{eq:Sfree_2d}, \ref{eq:Hfree_2d}) was used in two
dimensions.  These quantities for the concentric planar case were
obtained by a numerical inverse Laplace transform of
Eqs. (\ref{eq:tildeH_2d}, \ref{eq:tildeH_3d}), even so spectral
expansions can also be obtained via the residue theorem.

The decay time $T$ was estimated from the analysis of the logarithmic
derivative of the survival probability.  In fact, the long-time
relation (\ref{eq:S_expon}) implies that $-\dot{S}(t)/S(t) \approx
1/T$ over a broad range of times $t \in (t_1,t_2)$.  Here $t_1$ is the
timescale above which the long-time relation (\ref{eq:S_expon}) is
applicable, i.e., when the other terms of the spectral expansion
(\ref{eq:S_spectral}) can be neglected.  Strictly speaking, this
timescale is determined by the second eigenvalue of the diffusion
operator but in practice, it is sufficient to take $t_1$ to be of the
order of $T$ (e.g., $5T$).  The upper limit $t_2$, which formally
could be infinitely large, is necessary to eliminate statistical
uncertainties in the survival probability due to a limited number of
realizations.  In practical terms (see Appendix \ref{sec:estimation})
we choose the time interval $(t_1, t_2)$ in such a way that, except
for statistical uncertainties, $-\dot{S}(t)/S(t)$ remains
(approximately) constant.
Once the time range $(t_1, t_2)$ is set, the decay time can be
estimated as
\begin{equation}  \label{eq:T_MC}
T = \frac{t_2 - t_1}{\int_{t_1}^{t_2} dt ~ (-\dot{S}(t)/S(t))} \,, 
\end{equation} 
while the accuracy of this estimate can be measured by the norm of
fluctuations of $-\dot{S}(t)/S(t)$ around $1/T$:
\begin{equation}   \label{eq:deltaT_MC} 
\delta T = T^2 \sqrt{\frac{ \int_{t_1}^{t_2} dt [-\dot{S}(t)/S(t) - 1/T]^2 }{t_2 - t_1}} 
\end{equation} 
(see illustrations in Fig.~\ref{fig:slope} and further discussion in
Appendix \ref{sec:estimation}).

We emphasize that this estimation procedure is more informative than a
direct linear fit of $\ln S(t)$.  First, one can choose the
appropriate range $(t_1,t_2)$ and also evaluate the error $\delta T$.
Second, in cases when one particle has a much smaller diffusion
coefficient than the other particle, there may exist an intermediate
regime, in which the exponential factor $e^{-t/T}$ is affected by a
slowly varying prefactor $f(t)$ converging to a constant as
$t\to\infty$.  This prefactor may result in a systematic bias in the
estimated decay time $T$.  As such a bias is usually small, it is
difficult to appreciate from fitting $\ln S(t)$.  In turn, its effect
becomes more apparent when showing $-\dot{S}(t)/S(t)$.

We also estimated the MFET.  As the numerical simulations have been
performed with a time cut-off at $t_{\rm cut}$, one cannot compute
directly the MFET by taking an average over realizations $\T_i$ of the
first-encounter time.  Nevertheless, it can be estimated through other
quantities that are directly accessible.  The first one is the average
of the first-encounter times generated in each run, constrained to be
equal to $t_{\rm cut}$ when the particles have not yet met by the time
$t_{\rm cut}$, i.e.
\begin{equation}  \label{eq:Tcut}
\bar{\T} = \frac{1}{N} \sum_{i=1}^N  \min \{\T_i, t_{\rm cut}\},
\end{equation}
where $\T_i$ is the first-encounter time in the $i$-th realization if
there were no cut-off.  For large $N$, this empirical average
approximates the expectation
\begin{align}  \nonumber
\bar{\T} & \xrightarrow[N\to \infty]{}
\langle \min \{\T, t_{\rm cut}\} \rangle = \int\limits_0^{t_{\rm cut}} dt \, t\, H(t) + t_{\rm cut} \, \int\limits_{t_{\rm cut}}^\infty dt \, H(t) \\
\label{eq:barT_bound}
& = \int\limits_0^{t_{\rm cut}} dt \, S(t) = \langle \T \rangle - \int\limits_{t_{\rm cut}}^\infty dt \, S(t)
\end{align}
(here we omitted the arguments $\x_1,\x_2$ for brevity).  This
quantity is clearly a lower bound for the MFET $\langle \T \rangle$.
According to the second line, this estimate corresponds to the
truncation of the integral in Eq. (\ref{eq:MFPT}) at $t_{\rm cut}$.
If $t_{\rm cut} \gg T$, the long-time behavior of the survival
probability can be approximated as
\begin{equation}  \label{eq:Sapprox_tcut}
S(t) \simeq S(t_{\rm cut}) \, \exp(-(t-t_{\rm cut})/T),
\end{equation}
so that  
\begin{equation}
\langle \min \{\T, t_{\rm cut}\} \rangle \simeq \langle \T \rangle - T \, S(t_{\rm cut}).
\end{equation}
In this way, one can control the error of the estimate $\bar{\T}$ and
choose an appropriate $t_{\rm cut}$; in particular, $S(t_{\rm cut})$
should be very small.

The other manner to estimate the MFET is by computing the average with
the conditional probability density:
\begin{equation}
H_{\rm cond}(t) = H(t) \biggl(\int\limits_0^{t_{\rm cut}} dt' \, H(t')\biggr)^{-1} 
\end{equation}
(again, the dependence on $\x_1,\x_2$ is omitted here).  This density
is defined and well normalized for times from $0$ to $t_{\rm cut}$.
The corresponding conditional MFET reads then
\begin{equation}
\langle \T\rangle_{\rm cond} = \int\limits_0^{t_{\rm cut}} dt \, t \, H_{\rm cond}(t) .
\end{equation}
As $t_{\rm cut}$ goes to infinity, the conditional mean approaches
$\langle \T\rangle$.  Indeed, one gets  
\begin{align}  \nonumber
\langle \T\rangle_{\rm cond} & = \frac{\langle \T \rangle - t_{\rm cut} S(t_{\rm cut})
- \int\nolimits_{t_{\rm cut}}^\infty dt \, S(t)}{1 - S(t_{\rm cut})} \\
& \simeq  \frac{\langle \T \rangle - (t_{\rm cut} +T) S(t_{\rm cut})}{1 - S(t_{\rm cut})} \,,
\end{align}
where we used again the approximation (\ref{eq:Sapprox_tcut}) to get
the second relation.  One sees that $\langle \T\rangle_{\rm cond}$ is
very close to $\langle \T\rangle$ as soon as $t_{\rm cut} \gg T$.
From empirical data, the conditional MFET can be estimated as
\begin{equation} \label{eq:Tcut*}
\bar{\T}^{*} = \frac{\sum_{i=1}^N \T_i \, \I_{\T_i \le t_{\rm cut}}}{\sum_{i=1}^N \I_{ \T_i \le t_{\rm cut}} } \,,
\end{equation}
where $\I_{\T_i \le t_{\rm cut}} = 1$ if $\T_i \le t_{\rm cut}$, and
$0$ otherwise.  When $\bar{\T}$ and $\bar{\T}^{*}$ are close, they are
very good estimates of the MFET, as we only neglected some outlier
data (in all our simulations $S(t)$ is very small for $t=t_{\rm cut}$
and decays exponentially for $t>t_{\rm cut}$, which makes the weight
of those outliers negligible).

\section{Fixed target problem in 2D}
\label{Sec:Fixed_Target}

\begin{figure}
\includegraphics[width=0.45\textwidth]{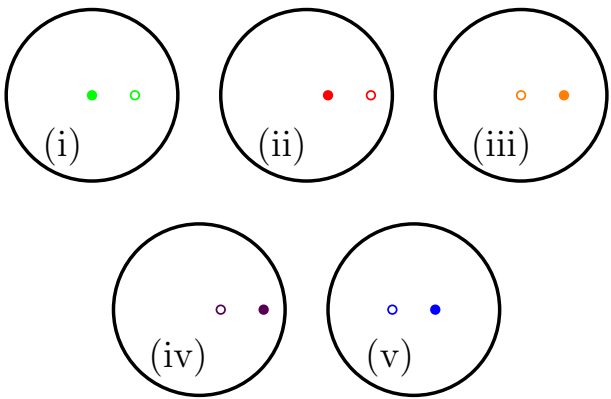} 
\begin{picture}(0,0)(0,0)
\Large
\put(-218,92){(i)}
\put(-137,92){(ii)}
\put(-56,92){(iii)}
\put(-178,12){(iv)}
\put(-96,12){(v)}
\end{picture}
\caption{
Five initial configurations with a diffusing particle (empty circle)
and a fixed target (filled circle) inside a disk of radius $R=10$.
The initial positions of the centers of the diffusing particle and of
the target are: (i) $\x_1 = (5,0)$ and $\x_2 = (0,0)$, (ii) $\x_1 =
(7.5,0)$ and $\x_2 = (2.5,0)$, (iii) $\x_1 = (0,0)$ and $\x_2 =
(5,0)$, (iv) $\x_1 = (2.5,0)$ and $\x_2 = (7.5,0)$, and (v) $\x_1 =
(-2.5,0)$ and $\x_2 = (2.5,0)$.}
\label{Fig:IC_Target}
\end{figure}

To gain intuition onto the dependence of the FET on the initial
positions, we start with the fixed target problem.  The comparison of
numerical results with available theoretical predictions will serve
for validating Monte Carlo simulations.  We consider the confining
domain $\Omega$ to be a disk of radius $R = 10$ with reflecting
boundary; a particle started from $\x_1$ diffuses with diffusion
coefficient $D_1 = 1/2$ towards an immobile target ($D_2 = 0$) fixed
at $\x_2$.  We fix the initial distance between the particles, $|\x_1
- \x_2| = 5$, and consider five configurations shown in
Fig. \ref{Fig:IC_Target}.

We will distinguish three regimes: short times ($t \lesssim t_B$) when
the boundary does not yet play any role; intermediate times ($t_B
\lesssim t \lesssim T$); and long times ($t\gtrsim T$), at which the
monoexponential decay of the survival probability is established.
Here $T$ is the decay time defined by \eqref{eq:Tx2}, whereas the time
scale $t_B$ will be defined below.

\begin{figure}
\includegraphics[width=0.45\textwidth]{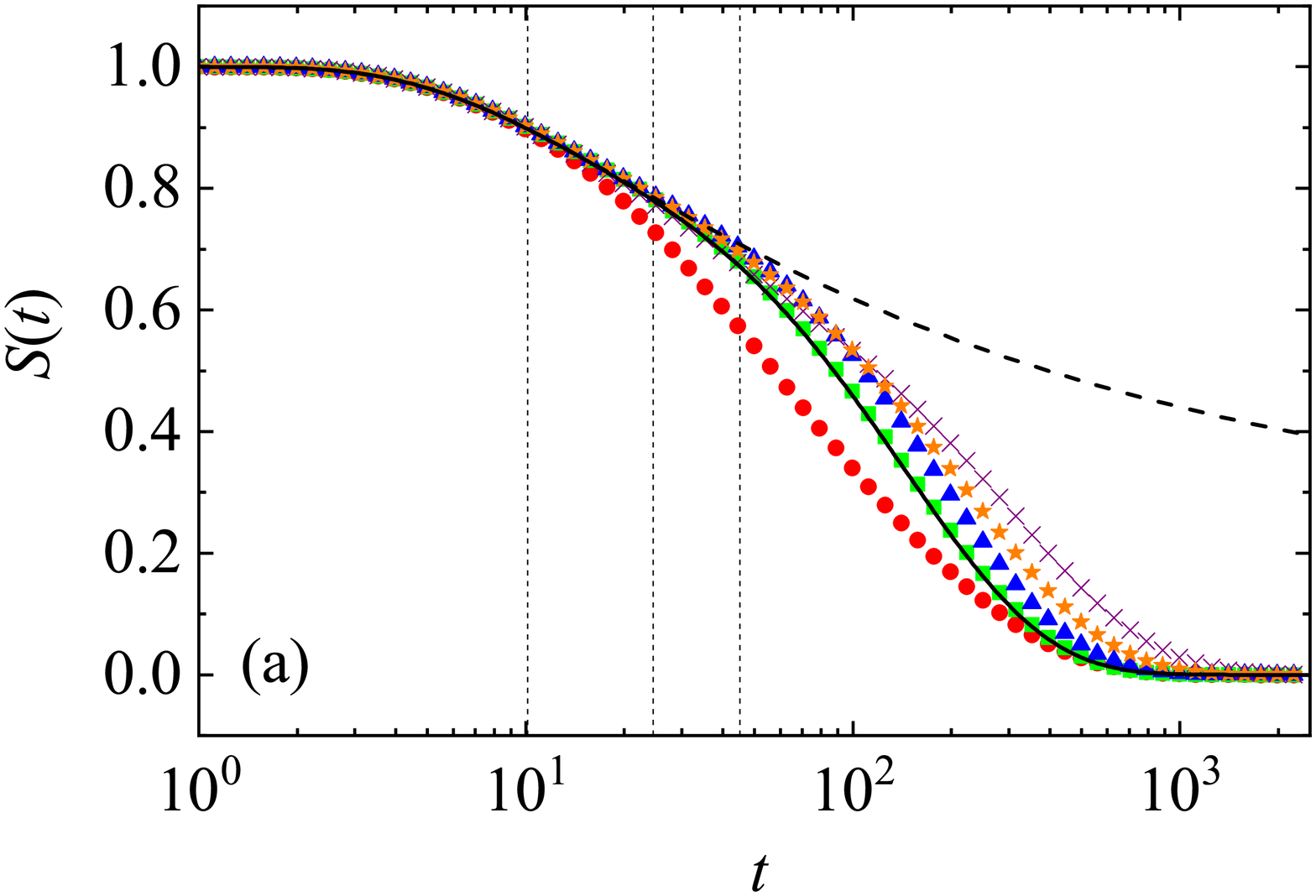} 
\includegraphics[width=0.45\textwidth]{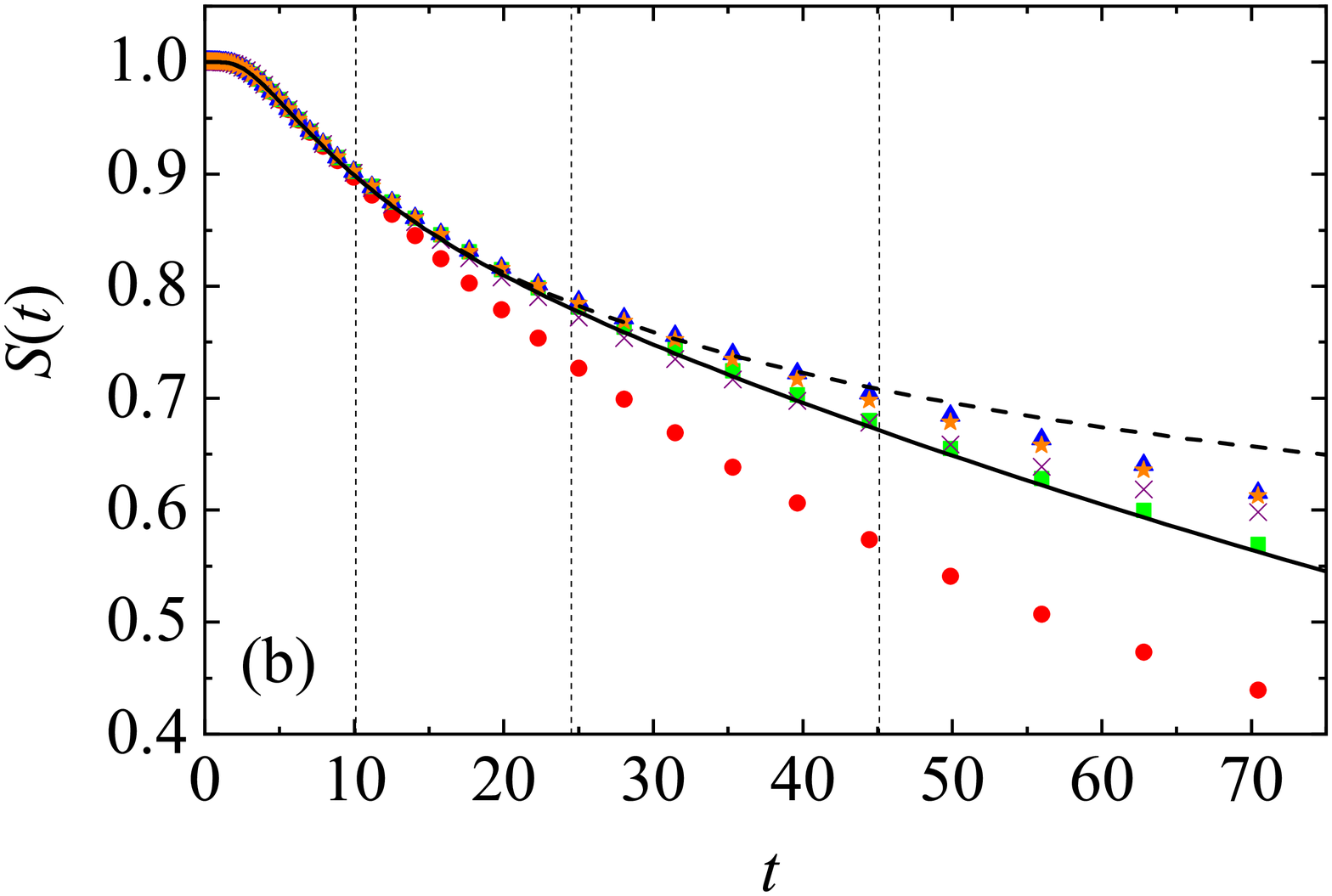} 
\includegraphics[width=0.45\textwidth]{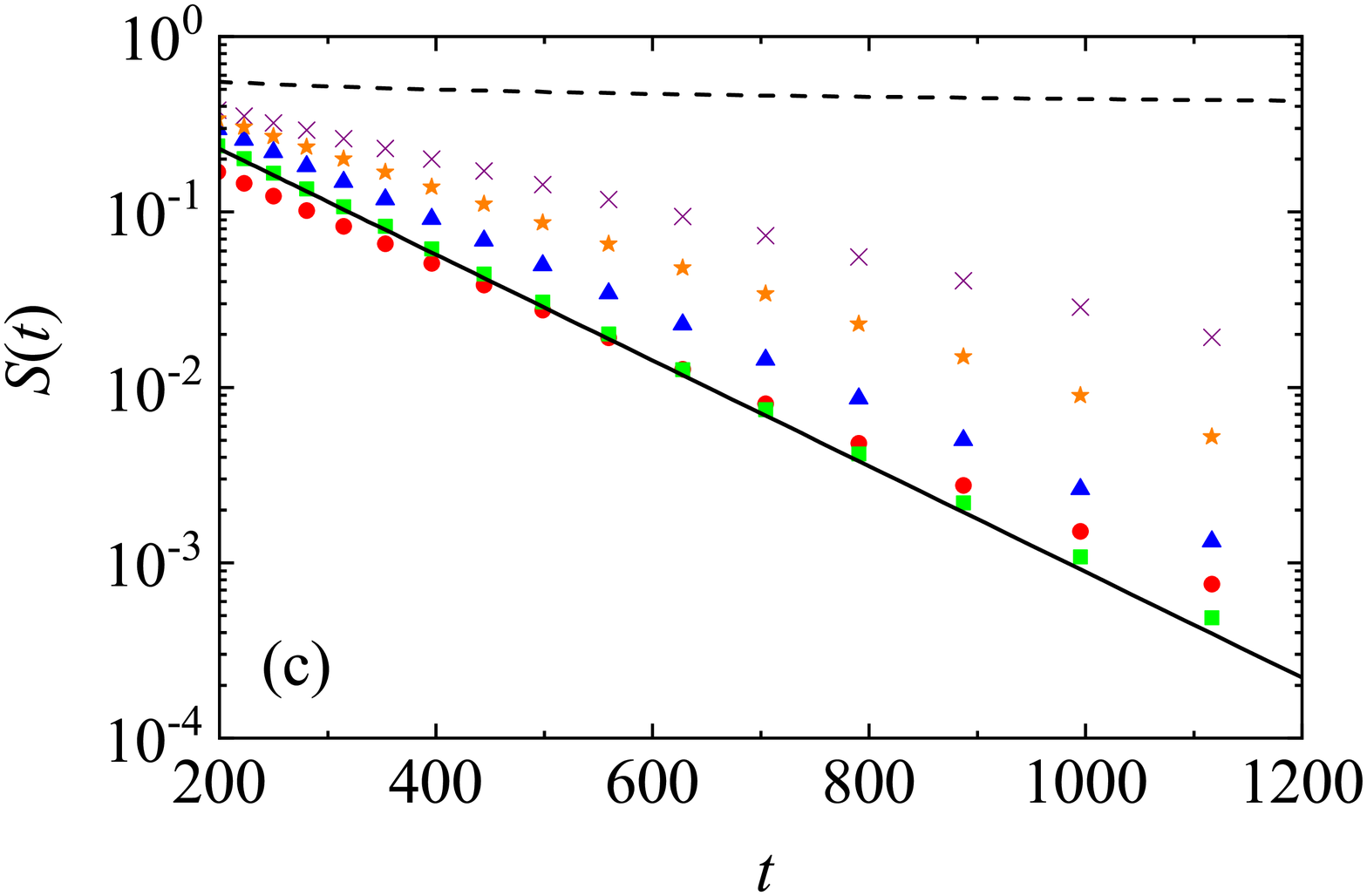} 
\caption{
Survival probability versus time for a particle of diffusion
coefficient $D_1=1/2$ in the search for a fixed target inside a disk
of radius $R = 10$.  Symbols present $S(t|\x_1,\x_2)$ for five
configurations of $\x_1$ and $\x_2$ described in
Fig.~\ref{Fig:IC_Target}: squares (i), circles (ii), stars (iii),
crosses (iv), and triangles (v).  Solid line is the exact solution for
case (i), obtained by numerical Laplace inversion of
Eqs. (\ref{eq:tildeS_23d}) and (\ref{eq:tildeH_2d}).  Dashed line
shows $S_{\rm free}(t|\x_1,\x_2)$ from Eq. (\ref{eq:Sfree_2d}).
Vertical dashed lines indicate the values of $t_B$: $10.1$ for cases
(ii) and (iv), $24.5$ for cases (i) and (iii), and $45.1$ for case
(v).  Note that $t_B$ corresponds to the time at which $S(t)$
separates from $S_\text{free}(t)$ for cases (i), (ii) and (v), but not
for cases (iii) and (iv).  Panels {\bf (a,b,c)} illustrate different
aspects of the same survival probabilities.  }
\label{Fig:SP_Target}
\end{figure}

Figure~\ref{Fig:SP_Target}(a) presents the survival probabilities for
five configurations shown in Fig. \ref{Fig:IC_Target}.  At short
times, the order in which $S(t)$ first deviates from $S_\text{free}$
is $S(\text{ii})\to S(\text{iv})\to S(\text{i})\to S(\text{iii})\to
S(\text{v})$, see Fig. \ref{Fig:SP_Target}(b).  The presence of the
reflecting boundary implies a reduction of the survival probability
with respect to the no boundary case (dashed line).  In fact,
confinement does not allow the diffusing particle to move too far away
from the target.  Then, in those initial arrangements where the
particles are closer to the boundary, they have more chances to meet
earlier.  Let us now introduce an empirical time scale $t_B$ to
describe when the boundary starts to matter,
\begin{equation}  \label{eq:tB}
t_B \equiv \frac{(L_M-\rho/2)^2}{2d D},
\end{equation}
where $L_M$ is the distance between the boundary and the middle point
of the initial positions of the particles (their centers).  With this
definition $t_B$(ii)$=t_B$(iv)$<t_B$(i)$=t_B$(iii)$<t_B$(v).

Equation (\ref{eq:tB}) has the drawback that it does not recognize
that the time at which the boundary starts to matter, is shorter for
case (ii) than for case (iv), and also shorter for case (i) than for
case (iii) (see Fig. \ref{Fig:SP_Target}(b)).  The reason for this
behavior is that for cases (ii) and (i) the diffusing particle starts
from the position that is closer to the boundary and it is therefore
likely for the diffusing particle to find the boundary rapidly and
then to move towards the target along the boundary \cite{Calandre14}.
In turn, if the target is closer to the boundary, the particle can get
farther both from the target and the boundary at short times.

The latter argument can be extended to explain the long-time behavior
of the survival probability.  For some locations of the target, there
could be extended regions in which the moving particle may diffuse for
a long time without approaching the target.  In particular, when the
target is centered, the survival probability at long times is expected
to be the smallest one, as confirmed by simulations.  In this
particular case, the sum of the distances from the starting point of
the moving particle to the target and to the boundary is constant,
i.e., it does not depend on the starting position of the moving
particle.

Figure~\ref{Fig:SP_Target}(c) illustrates the exponential decay
(\ref{eq:St_decay}) of the survival probability at long times, with
the decay time $T(\x_2)$ given by Eq. (\ref{eq:Tx2}).
Table~\ref{Tab:Target_MV} provides $T$ for the initial configurations
(i)-(v) described above.
The values of $T$ differ from each other, except for the cases (ii)
and (v), where $T \simeq 165$, highlighting the dependence of $T$ on
the position of the target but not on the initial position on the
diffusing particle.  Expectedly, the smallest $T$ is observed for the
centered target, while the configurations (iii) and (iv) yield larger
$T$ as the target is located far from the center of the disk (note
that a similar effect for the MFET $\langle \T \rangle$ was reported
in \cite{Condamin07b}).  The values of $T$ estimated from Monte Carlo
simulations are in excellent agreement with their theoretical
predictions from Eq. (\ref{eq:Tx2}).  We also stress that the decay
time $T$ is in very good agreement with its approximation by the
small-target asymptotic formula (\ref{eq:Tsmall_2d}), except for the
case (iv), in which the target is too close to the boundary, and
Eq. (\ref{eq:Tsmall_2d}) is not applicable.  We emphasize that the
second-order term in Eq. (\ref{eq:Tsmall_2d}) is significant: the
leading-order approximation (such as Eq. (\ref{eq:Tsmall})) would give
$T\approx 203$ for all initial configurations.

In Table~\ref{Tab:Target_MV} we also provide the values of two
estimates $\bar{\T}$ and $\bar{\T}^{*}$ of the MFET.  For the case
(i), Eq. (\ref{eq:MFPT_concentric}) yields the MFET $\langle \T
\rangle \approx 133$, which differs by only $2\%$ from both
estimates $\bar{\T}$ and $\bar{\T}^*$.
When comparing the cases (ii) and (v), one observes that their MFETs
are quite distinct, as opposed to almost identical values of $T$ in
these cases.  The initial configuration (ii) leads to a lower MFET
than (v) because the center of mass is closer to the boundary,
favoring the encounter of two particles at shorter times.  This
example illustrates the dependence of the MFET on the initial
position.  Besides, the estimates of the MFET are close to $T$ in the
cases (i) and (v).  In the case (v), the target is close to the center
of the disk (that avoids large void regions), and the relevance of the
boundary appears at larger times than in the other cases.  Notice
that, roughly, the following rule-of-thumb holds: the sooner $S(t)$
separates from $S_\text{free}$, the better the agreement between the
MFET and $T$ is.  This will also be seen to be case for two diffusing
particles (see Sec. \ref{sec:identical_long} and
\ref{Sec:Transition}).

\begin{table}
\centering
\begin{tabular}{ |c||c|c|c||c|c|c| }
 \hline
 & \multicolumn{3}{c||}{Decay time} & \multicolumn{3}{c|}{MFET} \\  \hline
 Case & $T_\text{num}$ &$T_{\rm asympt}$ & $T_\text{simu} (\delta T)$& $\bar{\T}_\text{num}$ & $\bar{\T}$ & $\bar{\T}^{*}$ \\
 \hline
(i)  & 144 &  152 &  146  (3.6) &  133  & 136 &  136 \\
(ii) & 165 &  174 &  167  (4.5) &  108  & 110 &  110 \\
(iii)& 218 &  231 &  219  (3.6) &  184  & 187 &  187 \\
(iv) & 305 &  558 &  308  (4.0) &  228  & 231 &  230 \\
(v)  & 165 &  174 &  167  (5.1) &  157  & 160 &  160 \\ \hline
\end{tabular}
\caption{
Several estimates of the decay time $T$ and MFET $\langle \T \rangle$
for the cases of Fig.~\ref{Fig:IC_Target}.  Here $T_\text{num} =
1/\lambda_1$ is obtained by means of the numerical computation of the
first eigenvalue $\lambda_1$ of the Laplace operator by a
finite-element method (implemented in the PDEtool, Matlab).  The
result for case (i) agrees with the exact value provided by
Eq. \eqref{eq:alpha_n}.  $T_{\rm asympt}$ is obtained by the
small-target asymptotic formula (\ref{eq:Tsmall_2d}), $T_\text{simu}$
is the value estimated from Monte Carlo simulations and $\delta T$ are
the corresponding errors obtained from Eq. (\ref{eq:deltaT_MC}).  Note
that the value of $T_{\rm asympt}$ for (iv) is too large because the
target is located near the boundary, and so Eq. (\ref{eq:Tsmall_2d})
is not applicable.
On the other hand, $\bar{\T}_{\rm num}$ is the estimate of $\langle \T
\rangle$ obtained by solving numerically the boundary value problem
Eqs.\eqref{eq:TavgEq}-\eqref{eq:TavginGamma} by a finite-element
method (FEM) implemented in PDEtool, Matlab.  Finally the two
estimates $\bar{\T}$ and $\bar{\T}^*$ of the MFET from
Eqs. (\ref{eq:Tcut}) and (\ref{eq:Tcut*}) are also given.  A minor but
systematic difference between $\bar{\T}_{\rm num}$ and these two
estimates can potentially be attributed to discretization effects in
both numerical methods (spatial discretization of FEM and temporal
discretization in Monte Carlo simulations).}
\label{Tab:Target_MV}
\end{table}

In summary, the survival probability in confinement changes,
especially at long times, if the initial positions of a diffusing
particle and a fixed target are swapped, unless the problem preserves
the symmetry after the swap (e.g. in the case (v)).

\begin{figure}[t]
\includegraphics[width=0.45\textwidth]{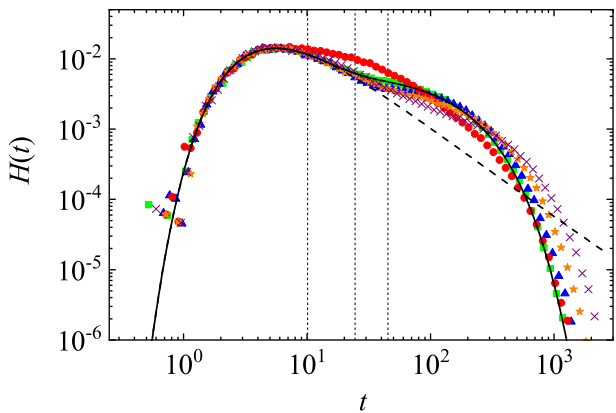} 
\caption{
FET probability density versus time for a particle of diffusion
constant $D_1=1/2$ in the search for a fixed target inside a disk of
radius $R=10$.  Symbols present simulation results for five
configurations shown in Fig. \ref{Fig:IC_Target}: squares (i), circles
(ii), stars (iii), crosses (iv), and triangles (v).  Solid line shows
the exact solution for the case (i), obtained by numerical Laplace
inversion of Eq. (\ref{eq:tildeH_2d}).  Dashed line is $H_{\rm
free}(t|\x_1,\x_2)$ from Eq. (\ref{eq:Hfree_2d}).  Vertical dashed
lines indicate the values of $t_B$: $10.1$ for cases (ii) and (iv),
$24.5$ for cases (i) and (iii), and $45.1$ for case (v).}
\label{Fig:Histo_Target}
\end{figure}

Figure~\ref{Fig:Histo_Target} illustrates the FET probability density
$H(t|\x_1,\x_2)$.  Let first note that the simulations for the case
(i) manifest an excellent agreement with theory.  One observes that
the densities coincide with the solution $H_{\rm free}(t|\x_1,\x_2)$
for the no boundary case at least until $t_B$.  At long times, the
densities exhibit an exponential decay, with the decay time $T$
depending the position of the target, as expected.

\section{Identical diffusing particles}
\label{Sec:Identical_Particles}

In this section, we will see what happens if the fixed target starts
to diffuse as the other particle.  In other words, we study the
statistics of the first-encounter time for two identical diffusing
particles with $D_1=D_2=D/2$, confined in a disk of radius $R$ with
reflecting boundary.

In the no boundary case, there is no difference between the problem
with a fixed target and the problem with two diffusing particles, as
the survival probability, given by Eq. (\ref{eq:Sfree_2d}), depends on
the sum of diffusion coefficients.  However, in the presence of a
reflecting boundary, these two problems are no longer equivalent and
we will compare them in this section.

\subsection{Two timescales}

First, we identify two timescales that control the behavior of the
survival probability: $t_F$, at which two particles {\it typically}
meet for the first time in the no boundary case, and $t_B$, above
which the influence of the boundary cannot be neglected.  The
timescale $t_F$ can be defined as the most probable FET, i.e., the
time at which the FET density $H_{\rm free}(t|\x_1,\x_2)$ is maximal
\cite{PartI}.  In three dimensions, taking the time derivative of the
explicit formula (\ref{eq:S_free3D}) and equating it to $0$ yield
\begin{equation}  \label{eq:tF}
t_F = \frac{(r - \rho)^2}{6 D} \,.
\end{equation}
In two dimensions, it was argued that Eq. (\ref{eq:tF}) still gives an
accurate estimate of the most probable FET \cite{Grebenkov18b}.  We
emphasize that the factor $6$ in the denominator does not depend on
the space dimensionality, given that the short-time asymptotic
behavior of the PDF is given by $e^{-(r-\rho)^2/(4Dt)}/t^{3/2}$ in all
dimensions.

The second timescale $t_B$ might naively be thought as being
determined by the initial distance from each particle to the boundary.
Such a distance would indeed determine a timescale for interaction of
a single particle to the boundary.  However, as we are interested in
the first-encounter time between two particles, the initial distances
between the particles and the boundary are less relevant.  In
contrast, if the particles are diametrically opposed and very close to
the boundary, the boundary starts to affect the motion of each
particle at very early times, but these times are not so relevant for
the first-encounter time, at least for small particles.  For this
reason, we keep using $t_B$ defined by Eq. (\ref{eq:tB}), as justified
below.

\begin{figure}[t]
\includegraphics[width=0.45\textwidth]{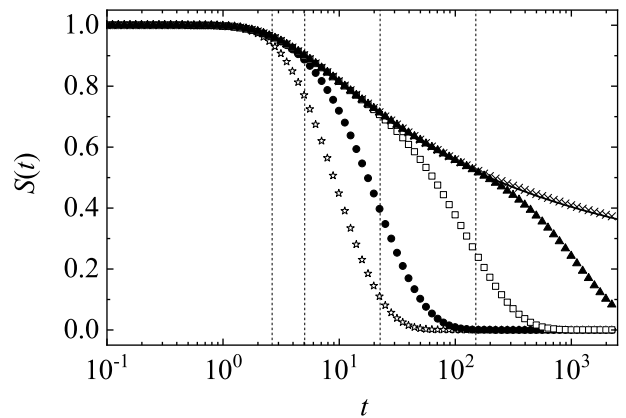} 
\caption{
Survival probability versus time for two diffusing particles with
equal diffusion coefficients $D_1=D_2=1/2$ that are initially placed
at $(-2.5,0)$ and $(0,2.5)$ inside a disk of radius $R$.  Symbols
present simulation results for $R=3.75$ (stars), $R=5$ (circles),
$R=10$ (squares), $R=25$ (triangles), and no boundary (crosses).
Solid line shows $S_{\rm free}(t|\x_1,\x_2)$ from
Eq. (\ref{eq:Sfree_2d}).  Vertical dashed lines indicate the values of
$t_B$: $2.6,~5.1,~22.6$, and $150$.  Here $t_F\simeq 2.7$.  }
\label{Fig:SP_Identical_DifR}
\end{figure}

Now we can study how the survival probability depends on the size of
the domain and on the initial positions of the particles.  We first
plot in Fig.~\ref{Fig:SP_Identical_DifR} the survival probability for
two particles initially placed at $\x_1=(-2.5,0)$ and $\x_2=(0,2.5)$
for different values of the domain radius $R$.  As expected, all
simulation results coincide with $S_{\rm free}(t|\x_1,\x_2)$ until
$\sim t_B$ corresponding to each value of $R$.  It is also observed
that the survival probability decays faster for lower $R$.

\begin{figure}[t]
\includegraphics[width=0.40\textwidth]{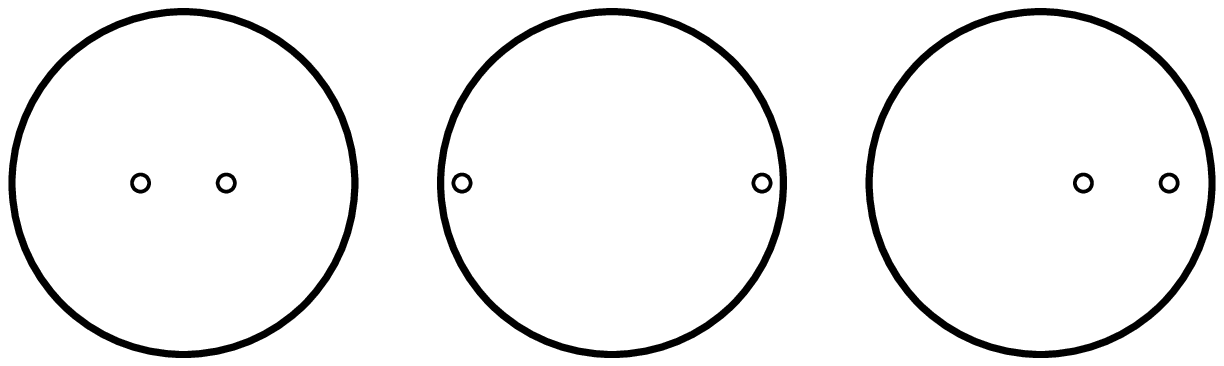} 
\vskip 3mm
\includegraphics[width=0.45\textwidth]{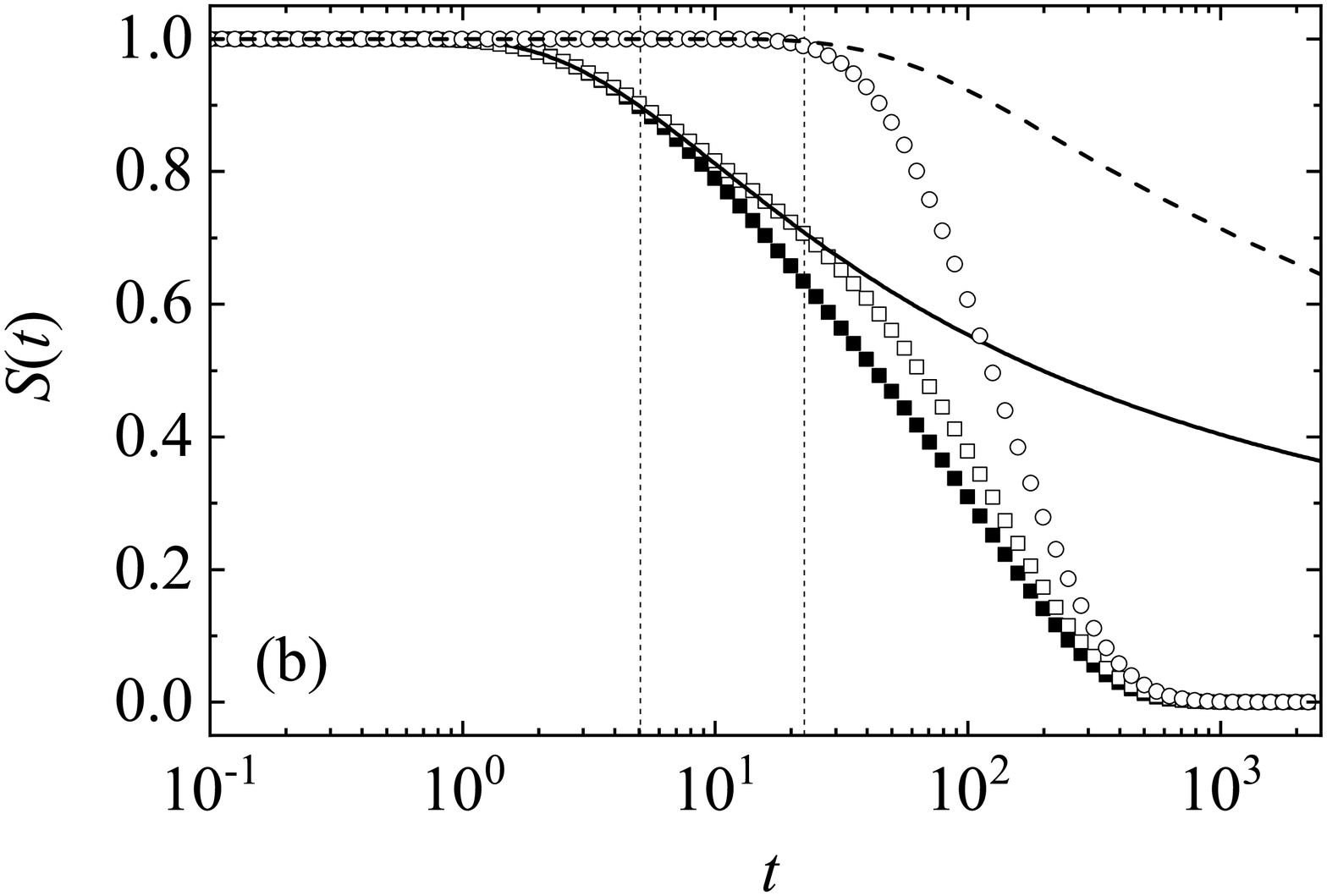} 
\includegraphics[width=0.45\textwidth]{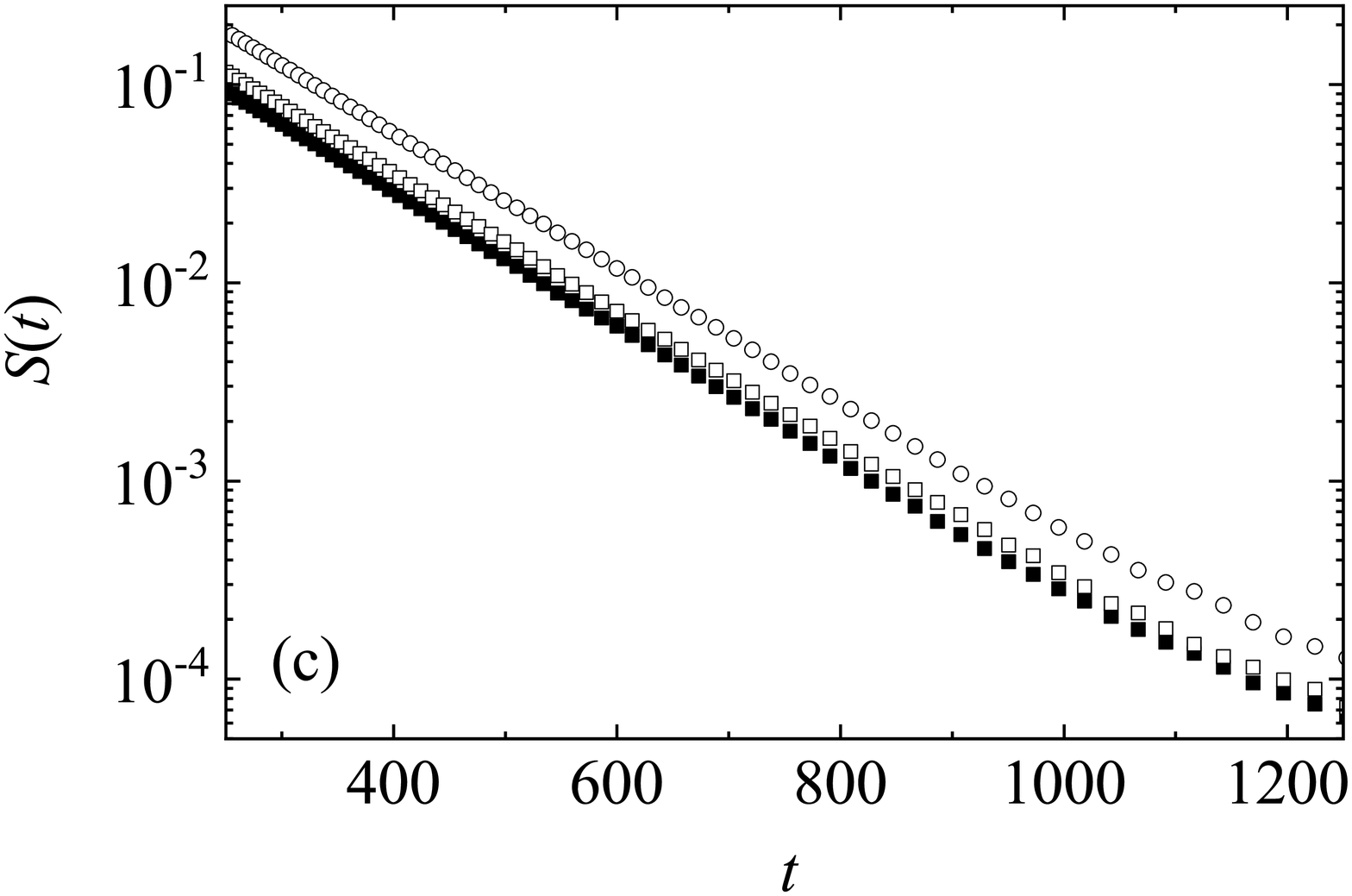} 
\begin{picture}(0,0)(0,0)
\large
\put(-195,325){(i)}
\put(-125,325){(ii)}
\put(-53,325){(iii)}
\end{picture}
\caption{
{\bf (a)} Three initial configuration of two diffusing particles with
equal diffusion coefficients $D_1 = D_2 = 1/2$ inside a disk of radius
$R=10$, with the initial positions: (i) $(-2.5,0)$ and $(2.5,0)$, (ii)
$(-8.75,0)$ and $(8.75,0)$, and (iii) $(0,2.5)$ and $(0,7.5)$.  {\bf
(b)} Survival probability versus time over a broad range of times
(logarithmic scale for horizontal axis); {\bf (c)} Long-time behavior
(logarithmic scale for vertical axis).  Symbols refer to the above
configurations: (i) empty squares, (ii) circles, and (iii) filled
squared.  Lines represent $S_{\rm free}(t|\x_1,\x_2)$ from
Eq. (\ref{eq:Sfree_2d}) for the initial inter-particle distances $r =
5$ (solid) and $r = 17.5$ (dashed).  Vertical dashed lines indicate
the values of $t_B=5.1$ [case (iii)] and $t_B=22.6$ [cases (i) and
(ii)].}
\label{Fig:SP_Identical_D1}
\end{figure}

Next, Fig.~\ref{Fig:SP_Identical_D1}(a) shows the survival probability
for three initial configurations with fixed $R=10$.  In configurations
(i) and (ii), the center of mass of the two particles is at the
origin, implying the same time $t_B \simeq 22.6$ according to
Eq. (\ref{eq:tB}).  One observes that the deviation from the no
boundary case occurs around this time, even though the two particles
are much closer to the boundary in the case (ii).  The survival
probability at $t_B$ is smaller in the case (i).  In turn, in
configurations (i) and (iii), the initial distances between the
centers of the particles is the same, but both particles are shifted
towards the boundary in the case (iii).  The corresponding survival
probabilities are different, highlighting their dependence on the
initial positions of both particles (not only on their initial
distance, as in the no boundary case).  In particular, the simulation
results deviate from $S_{\rm free}(t|\x_1,\x_2)$ with $r=5$ around
$t_B \simeq 5.1$.  We conclude that the center of mass is a useful
indicator of the time scale $t_B$ at which the survival probability
starts to differ from its counterpart in the no boundary case.

\subsection{Long-time decay}
\label{sec:identical_long}

While the above discussion focused on the short-time behavior, we now
study the survival probability at long times: $t\gg \max \{ t_B,
t_F\}$.  As the confining domain $\Omega$ is bounded, the survival
probability exhibits an exponential decay (\ref{eq:S_expon}).  We
estimate the decay time $T$ from Fig.~\ref{Fig:SP_Identical_D1}(b) and
analyze the dependence of $T$ on the parameters.  For three initial
configurations, the numerical points follow parallel straight lines,
while their linear fit yields the same decay time $T = 127 \pm 5$ (see
Table~\ref{Tab:Identical_MV}).  In fact, the initial condition appears
only in the prefactor in Eq. (\ref{eq:S_expon}), which shifts the
curves vertically.
In other words, at long times, the system almost forgets about the
initial condition, in contrast to the case of a fixed target, where
$T$ varied with the position of the target.

\begin{figure}[t]
\includegraphics[width=0.45\textwidth]{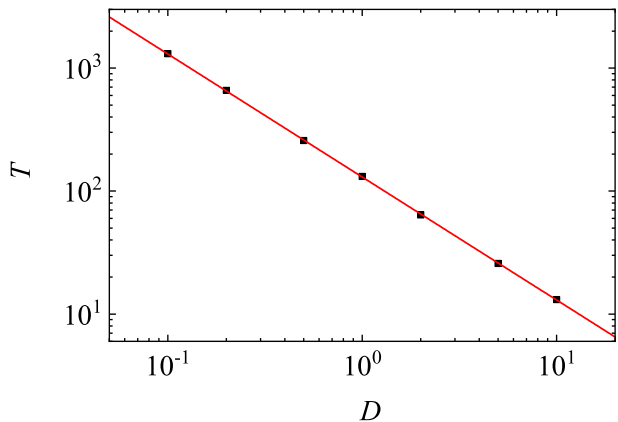}  
\caption{
The decay time $T$ versus $D$ for two diffusing particles with $D_1 =
D_2 = D/2$ placed initially at $(0,0)$ and $(2.5,0)$ in a disk of
radius $R=10$.  Squares represent simulation results, while solid line
is a linear fit with slope $-1$. }
\label{Fig:T_VS_Size2D}
\end{figure}

\begin{figure}[t]
\includegraphics[width=0.45\textwidth]{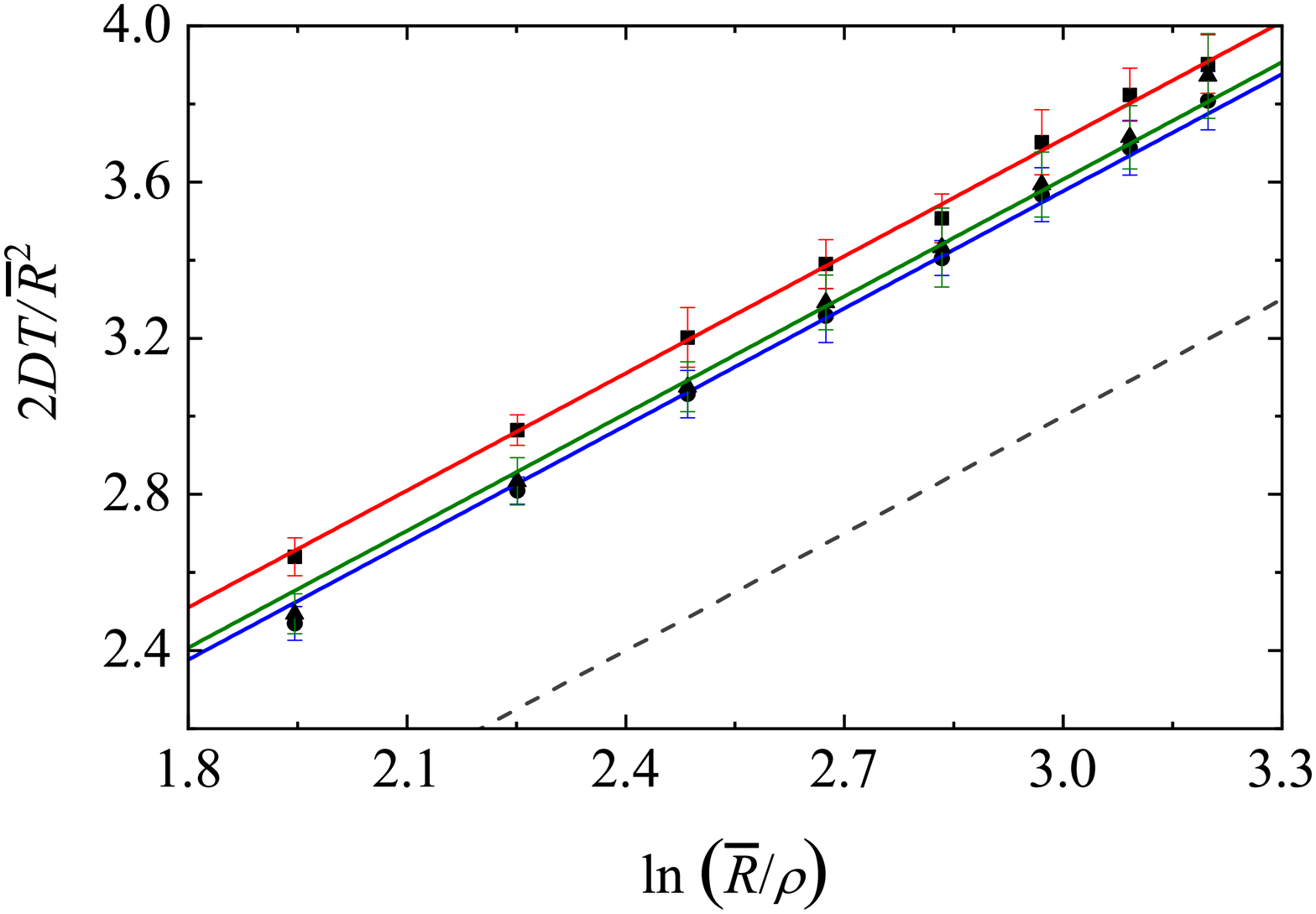} 
\caption{
Scaled decay time $2DT/\bar{R}^2$ versus $\ln(\bar{R}/\rho)$ for two
diffusing particles inside a disk of radius $R$, ranging from $7.5$ to
$25$.  Symbols show the simulation results for $D_1=0.9$ and $D_2=0.1$
(squares), $D_1=D_2=0.5$ (circles), and $D_1=1$ and $D_2=0.5$
(triangles).  Particle 2 (the one with the smallest diffusion
coefficient) is initially placed at the center whereas particle 1 is
placed at $(5,0)$.  The solid lines correspond to
Eq.~\eqref{eq:Tsmall_diff2} with $C_2=1$ and, from top to bottom,
$A(0.9/0.1)=0.71$, $A(1/0.5)=0.61$, and $A(0.5/0.5)=0.58$,
respectively.  As a reference, the dashed line represents the case
with no correction term ($A=0$). }
\label{Fig:T_VS_Size2D_2}
\end{figure}

In the small target limit, $\rho \ll \bar{R}$, it is instructive to
check whether the asymptotic formula (\ref{eq:Tsmall}), derived in the
case of a fixed target, is valid for two diffusing particles with $D =
D_1 + D_2 = 2D_1$:
\begin{equation}  \label{eq:Tsmall_diff}
T \simeq \frac{\bar{R}^2}{2D} \ln (\bar{R}/\rho)  \qquad (\rho \ll \bar{R}).
\end{equation}
A similar claim for the MFET was recently proved in the
three-dimensional case \cite{Lawley19}.
First, we observe in Fig.~\ref{Fig:T_VS_Size2D} that $T$ is indeed
proportional to $1/D$.  Each point corresponds to a different value of
the diffusion coefficient $D=2D_1$.  A linear fit in the double
logarithmic scale yields the expected slope of $-1$.  Second, we
analyze in Fig.~\ref{Fig:T_VS_Size2D_2} how $T$ changes with the size
$\bar{R}$ of the confining domain.  We find that our simulation
results are well described by the formula
\begin{equation}  \label{eq:Tsmall_diff2} 
T \simeq \frac{\bar{R}^2}{2D} \biggl(C_2  \ln (\bar{R}/\rho) + A(D_1/D_2) + \ldots \biggr),
\end{equation}
where $C_2=1$, $A(D_1/D_2)$ is a dimensionless function of $D_1/D_2$,
and $\ldots$ refers to next-order corrections, which are small for
$\rho \ll \bar{R}$ and not accessible from our simulations.  Even
though this section was focused on identical particles with $D_1 =
D_2$, i.e., only one value $A(1)$, we keep the general form
$A(D_1/D_2)$ that will be discussed for $D_1/D_2 \ne 1$ in
Sec. \ref{Sec:Transition}.  This means that the asymptotic formula
(\ref{eq:Tsmall}) for a fixed target reproduces the main logarithmic
term for the case of two diffusing particles.  Expectedly, the leading
term in Eq. \eqref{eq:Tsmall_diff2} with $D= 2 D_1$ is twice smaller
than that in Eq. \eqref{eq:Tsmall} with $D=D_1$, i.e., the decay is
faster in the present case of two identical searchers.  In other
words, to obtain the same asymptotic decay for a fixed target, the
searcher would need to have the twice larger diffusivity.  We also
outline that the leading (logarithmic) term in
Eq. \eqref{eq:Tsmall_diff2} is inaccurate due to the existence of the
$O(1)$ correction term $A(D_1/D_2)$, as confirmed by our simulations.
Getting a rigorous derivation of Eq. \eqref{eq:Tsmall_diff2} and
finding the correction term $A(D_1/D_2)$ present an interesting open
problem.  Note that other properties of the decay time, such as its
dependence on the number of searchers, were investigated in
\cite{Nayak20}.

\subsection{MFET}

Another important quantity is the MFET defined in Eq. (\ref{eq:MFPT}).
Table~\ref{Tab:Identical_MV} provides the values of two estimates
$\bar{\T}$ and $\bar{\T}^{*}$ of the MFET for the three initial
configurations shown in Fig.~\ref{Fig:SP_Identical_D1}(a).  In
contrast to the decay time $T$, the MFET depends on the initial
positions of the particles.  Interestingly, the MFET can be either
smaller, or larger than $T$ (recall that in the case of a fixed
target, we always observed that the MFET is smaller than $T$,
cf. Table \ref{Tab:Target_MV}).  In the small target limit, the main
contribution to the MFET comes from long trajectories that explore the
whole confining domain and correspond to the exponential decay of the
survival probability.  In this limit, the MFET is typically of the
order of $T$, while its variations can be caused by the prefactor in
Eq. (\ref{eq:S_expon}) which depends on the initial positions of both
particles.  One can observe a clear correlation between this prefactor
(that shifts the curves in Fig.~\ref{Fig:SP_Identical_D1} (b)) and the
values of the MFET in Table~\ref{Tab:Identical_MV}.

\begin{table}
\centering
\begin{tabular}{ |c|c|c|c| }  
 \hline
  & Decay time & \multicolumn{2}{c|}{MFET} \\  
 \hline
 Case & $T$ ($\delta T$)& $\bar{\T}$ & $\bar{\T}^{*}$ \\
 \hline
(i)   & 126 (2.4) & 107 &  107  \\
(ii)  & 127 (2.9) & 162 &  162  \\
(iii) & 127 (4.9) &  91 &   91  \\  \hline
\end{tabular}
\caption{
The decay time $T$, estimated error $\delta T $, and two estimates
$\bar{\T}$ and $\bar{\T}^{*}$ of the MFET from Eqs. (\ref{eq:Tcut},
\ref{eq:Tcut*}), for two diffusing particles with $D_1=D_2=1/2$ inside
a disk of radius $R=10$.  The initial positions of the particles are:
(i) $\x_1=(-2.5, 0)$ and $\x_2=(2.5, 0)$; (ii) $\x_1=(-8.75,0)$ and
$\x_2=(8.75, 0)$; (iii) $\x_1=(2.5,0)$ and $\x_2=(7.5,0)$, see
Fig.~\ref{Fig:SP_Identical_D1}.  Note that Eq. (\ref{eq:Tsmall_diff})
underestimates the decay time as $T \approx 101.6$, whereas the
inclusion of the correction term $A(1) \approx 0.58$ in
Eq. (\ref{eq:Tsmall_diff2}) gives $T\approx 127.8$, in perfect
agreement with the Monte Carlo estimate. }
\label{Tab:Identical_MV}
\end{table}

\subsection{Probability density}

To further highlight the relevance of the boundary, we study the shape
of the FET probability density $H(t|\x_1,\x_2)$.  In the no boundary
case, this density has a single hump around $t_F$: as $t$ grows, the
probability of first encounter initially increases (as both particles
need to travel a minimum distance to meet), and then slowly decreases
(as particles can diffuse too far away from each other).  The
extremely slow decay (\ref{eq:Hfree_2d_decay}) of $H_{\rm
free}(t|\x_1,\x_2)$ leads to infinite MFET.

The reflecting boundary changes completely this long-time behavior,
given that $H(t|\x_1,\x_2)$ exhibits an exponential decay inherited
from Eq. (\ref{eq:S_expon}).  In fact, the boundary prevents diffusing
particles from moving far away from each other, thereby eliminating
too long trajectories that were possible in the no boundary case.

\begin{figure}[t]
\includegraphics[width=0.45\textwidth]{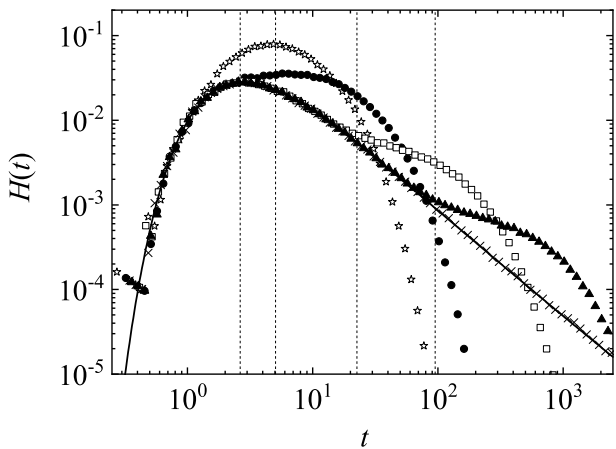} 
\caption{
FET probability density versus time for two diffusing particles of
diffusion constants $D_1 = D_2 = 1/2$, initially placed at $(-2.5,0)$
and $(0,2.5)$ inside a disk of radius $R$.  Symbols represent
simulation results for $R=3.75$ (stars), $R=5$ (circles), $R=10$
(squares), $R=25$ (triangles), and $R = \infty$ (crosses).  Solid line
shows $H_{\rm free}(t|\x_1,\x_2)$ from Eq. (\ref{eq:Hfree_2d}).
Vertical dashed lines indicate the values of $t_B$: $2.6,~5.1,~22.6$,
and $150$.}
\label{Fig:Histo_Identical_DifR}
\end{figure}

In Fig.~\ref{Fig:Histo_Identical_DifR} we show the FET probability
density for two particles, whose centers were initially placed at
$(-2.5,0)$ and $(2.5,0)$, with several values of the boundary radius
$R = 3.75,~5, ~10,$ and $25$.  Hence, in these cases, $t_B=
2.6,~5.1,~22.6$, and $150$, respectively, but $t_F = 2.7$ is the same.
As $t_B$ increases, the FET probability density coincides with $H_{\rm
free}(t|\x_1,\x_2)$ over a broader range of times $t < t_B$ and thus
widens.  When $t_B \gg t_F$, one observes the emergence of a second
hump around $t_B$.

\begin{figure}[t]
\includegraphics[width=0.45\textwidth]{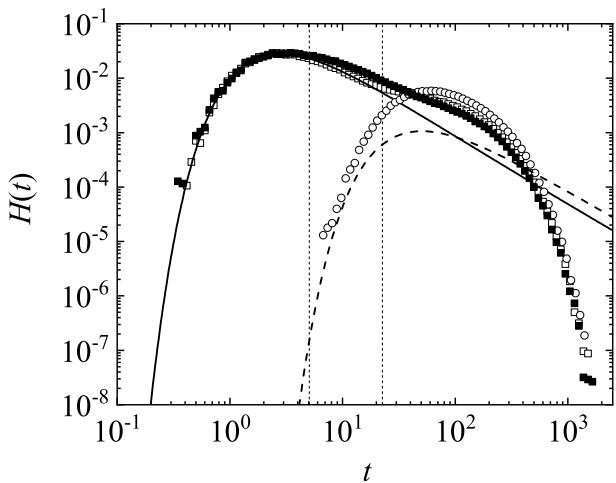} 
\caption{
FET probability density versus time for two diffusing particles of
diffusion constants $D_1 = D_2 = 1/2$ inside a disk of radius $R=10$,
with the initial positions of the particles: (i) $(-2.5,0)$ and
$(2.5,0)$ (empty squares), (ii) $(-8.75,0)$ and $(8.75,0)$ (circles),
and (iii) $(0,2.5)$ and $(0,7.5)$ (filled squares), see
Fig.~\ref{Fig:SP_Identical_D1}.  Lines represent $H_{\rm
free}(t|\x_1,\x_2)$ from Eq. (\ref{eq:Hfree_2d}) with the initial
inter-particle distances $r = 5$ (solid) and $r = 17.5$ (dashed).
Vertical dashed lines indicate the values of $t_B=5.1$ [case (iii)]
and $t_B=22.6$ [cases (i) and (ii)].}
\label{Fig:Histo_Identical_D1}
\end{figure}

Similar arguments can be used to describe
Fig.~\ref{Fig:Histo_Identical_D1} that shows the FET probability
density for three different configurations of particles in the same
bounded domain with $R=10$ (as illustrated in
Fig. \ref{Fig:SP_Identical_D1}(a)).  In cases (i) and (iii), the
inter-particle distance $r = 5$ is the same, and two probability
densities are close to each other (with the maximum around the same
$t_F\simeq 2.7$), even so they start deviating from $H_{\rm
free}(t|\x_1,\x_2)$ at different times $t_B$.  In turn, the case (ii)
with a larger distance $r = 17.5$ has larger $t_F\simeq 45.4$ so that
the maximum of the FET probability density is shifted toward longer
times.  As $t_B\simeq 22.6$ here is smaller than $t_F$, the FET
density exhibits considerable deviations from $H_{\rm
free}(t|\x_1,\x_2)$ over the relevant range of times.  Finally, all
the densities are very close to each other at long times, given that
the decay time $T$ does not depend on the initial positions of the
particles.

\section{Transition from immobile to mobile target}
\label{Sec:Transition}

In Sec.~\ref{Sec:Fixed_Target} and~Sec.~\ref{Sec:Identical_Particles},
we studied separately two scenarios of bimolecular reactions: a
diffusing particle searching for a fixed target, and two identical
diffusing particles searching to meet one another.  These scenarios
exhibited different properties because the fixed target introduced a
memory on the initial condition that affects the behavior of the
survival probability both at short and long times.  Here, we consider
particles with different diffusion coefficients to study transition
between two scenarios.  In fact, as the diffusion coefficient $D_2$
stands in front of the Laplace operator in
Eq. (\ref{eq:Sgeneral_def}), the limit $D_2 \to 0$, corresponding to a
fixed target scenario, is singular.  This is the mathematical origin
of distinct behaviors of the survival probability in the above two
scenarios.  In physical terms, the timescale associated with the
motion of the second particle, $L^2/D_2$, is infinite at $D_2 = 0$
(here $L$ is an appropriate length scale, e.g., $L = \bar{R}$).  In
turn, if $D_2$ is small (as compared to $D_1$) but strictly positive,
one can expect that the survival probability behaves at times $t \ll
L^2/D_2$ as in the case of a fixed target, and then switches to the
behavior for two mobile particles at longer times $t \gtrsim L^2/D_2$.
In other words, a smooth transition between two scenarios can be
expected.

\begin{figure}[!t]
\includegraphics[width=0.45\textwidth]{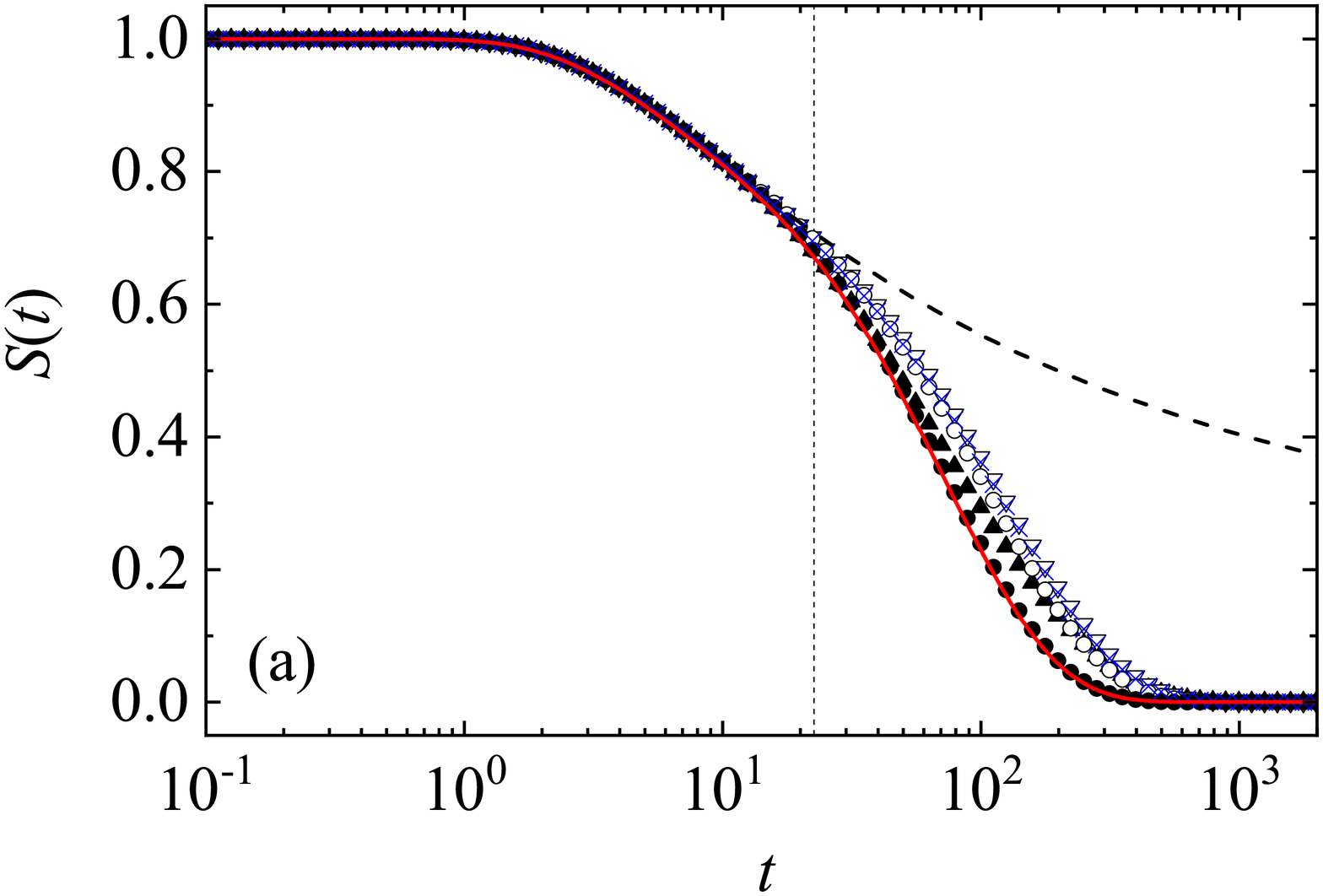} 
\includegraphics[width=0.45\textwidth]{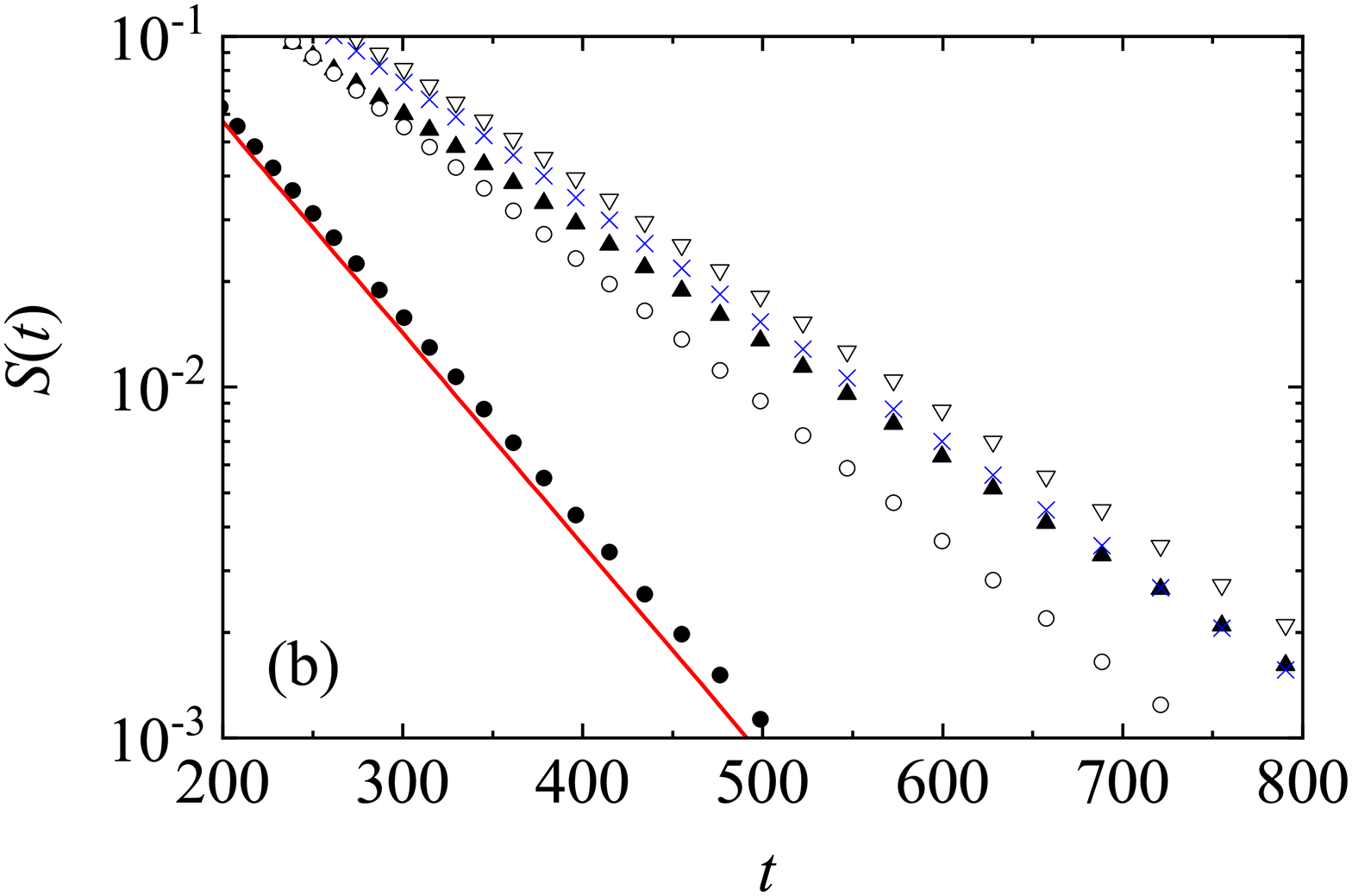} 
\includegraphics[width=0.45\textwidth]{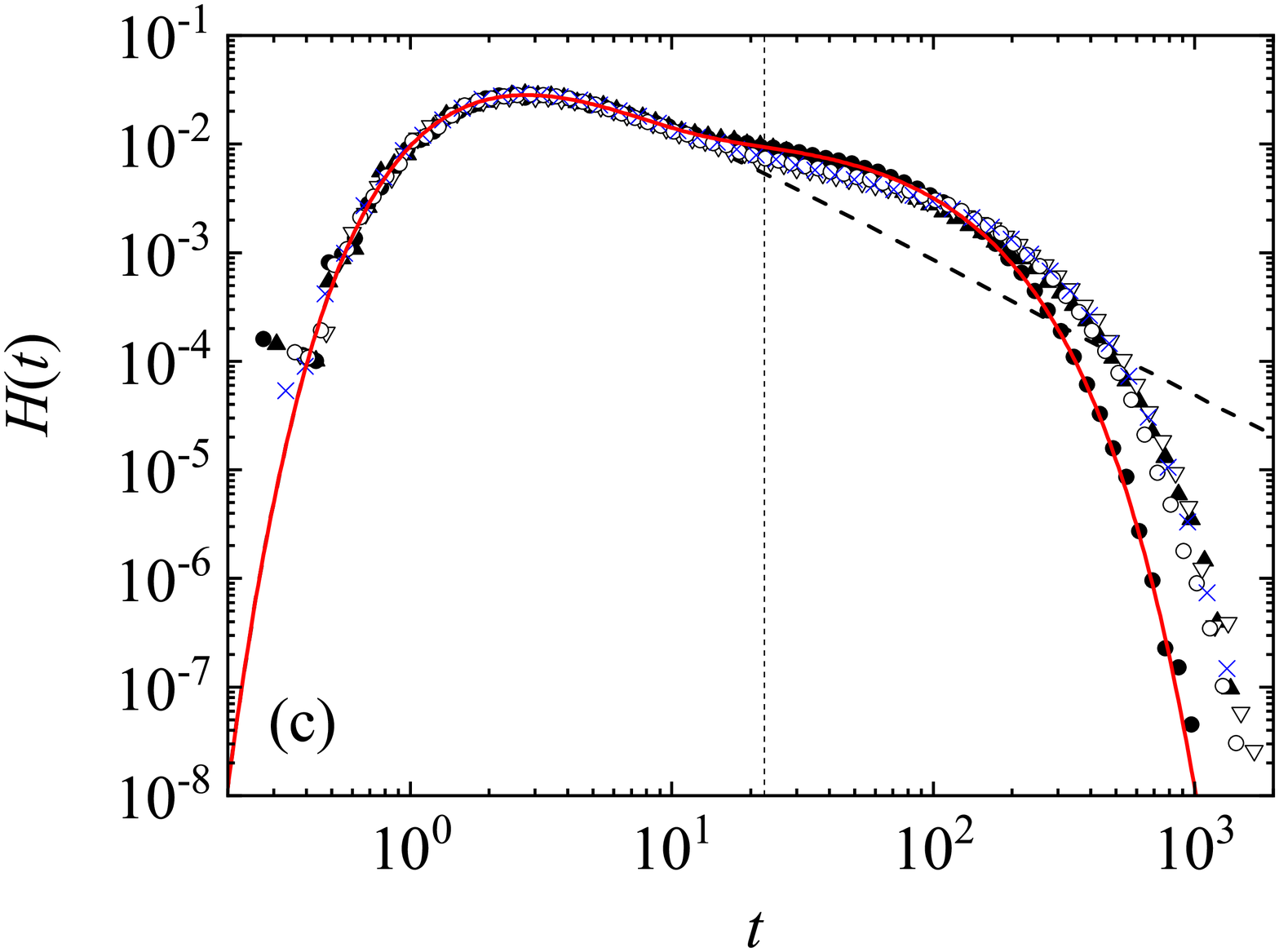} 
\caption{
{\bf (a,b)} Survival probability for two diffusing particles with: (i)
$\{D_1, D_2 \}=\{0,~1\}$ (filled circles), (ii) $\{1/10,~9/10\}$
(filled triangles), (iii) $\{1/2,~1/2\}$ (crosses), (iv)
$\{9/10,~1/10\}$ (empty triangles), and (v) $\{1,~0\}$ (empty
circles).  In the initial state, the particle with diffusion constant
$D_1$ is located at $(0,0)$ and the other is placed at $(5,0)$ inside
a disk of radius $R = 10$.  Solid line is the exact solution for the
case $\{D_1, D_2 \}=\{0,~1\}$ and dashed line presents $S_{\rm
free}(t|\x_1,\x_2)$.  Vertical dashed line indicates the value of $t_B
= 22.6$.  {\bf (c)} FET probability density for the same
configurations.  Short-time deviations are caused by the binning
artifact and a limited number of realizations with small FET. }
\label{Fig:SP_DifD}
\end{figure}

To clarify this transition, we run simulations for particles with
different $D_1$ and $D_2$ such that $D=D_1+D_2=1$ is fixed.  The first
particle is located at the center of a disk of radius $R=10$ and the
second one is at a distance $r = 2.5$.  The survival probability is
shown in Fig.~\ref{Fig:SP_DifD}(a) for five cases: (i) $D_1 = 0$, $D_2
= 1$, (ii) $D_1 = 0.1$, $D_2 = 0.9$, (iii) $D_1 = D_2 = 0.5$, (iv)
$D_1 = 0.9$, $D_2 = 0.1$, and (v) $D_1 = 1$, $D_2 = 0$.  In this
setting, $t_B \simeq 22.6$ for all cases so that the survival
probabilities remain close to $S_{\rm free}(t|\x_1,\x_2)$ for
$t\lesssim t_B$.  Afterwards, the curves start to deviate from each
other, showing that the survival probability depends explicitly on
$D_1$ and $D_2$, and not only on their sum.

The long-time behavior of the survival probability is detailed in
Fig.~\ref{Fig:SP_DifD}(b).  If one of the particles is fixed (cases
(i) and (v)), the survival probability exhibits a faster decay as
compared to the cases (ii)-(iv) when both particles diffuse.  
One sees that when the sum of the diffusion coefficients is fixed,
setting one of them equal to zero seems to be detrimental to the
survival probability at long times.  This statement can be called the
``anti-Pascal principle'', by opposition to the ``Pascal principle''.
The latter states that the survival probability of a mobile target is
less than or equal to the survival probability of an immobile target
when the diffusion coefficient of the moving particle is fixed
\cite{Moreau03}.  In other words, if the diffusion coefficient of a
``hunter'' is fixed, an immobile ``prey'' has more chances to survive
than a mobile one.  However, when the sum of diffusion coefficients is
fixed, the motion of the ``hunter'' is slower if the ``prey'' also
diffuses, and thus the mobile ``prey'' survives longer.  The fastest
decay corresponds to the case (i) when the fixed target is located at
the center of the disk because it is the most accessible for the
diffusing particle, implying faster encounters.

While the decay time is clearly different for cases (i) and (v) with a
fixed target, the long-time behavior of the survival probability in
cases (ii)-(iv) is rather similar.  In fact, according to
Eq. (\ref{eq:S_expon}), the decay rate $T$ is independent of the
starting positions, i.e., it should be the same for cases (ii) and
(iv).  This is confirmed by our simulations (see also the estimated
values in Table~\ref{Tab:DifD_MV}).  In turn, the decay time in the
case (iii) of equal diffusivities is by $4\%$ smaller than in cases
(ii) and (iv).  We note, however, that such a small difference could
still be an artifact of numerical simulations or of an estimation
procedure from the datapoints, for which the monoexponential decay may
not be fully established at the available time scales.

Another important point is the prefactor, which is responsible of the
weak dependence of the long-time exponential decay of the survival
probability on the initial condition, as in the case of identical
particles.  This prefactor can lead to different MFETs, depending on
whether the particle is close to the boundary or not.  Two estimates
of the MFET are provided in Table~\ref{Tab:DifD_MV}.  One observes
that encounters are faster when the particle with the larger diffusion
coefficient is close to the boundary.

Similarly, the FET probability densities are also close to each other
(Fig.~\ref{Fig:SP_DifD}(c)).  In all considered cases, the probability
densities exhibit a single maximum around $t_F \simeq 4$.
Interestingly, at times $t \gtrsim t_B$, the presence of the
reflecting boundary shifts the probability densities upwards, as
compared to the free case (dashed line).  A visual inspection suggests
the possible presence of inflection point(s) for the curve
$H(t|\x_1,\x_2)$.

\begin{table}
\centering
\begin{tabular}{ |c|c|c|c|c|c| } 
 \hline
  \multicolumn{3}{|c|}{} & Decay time & \multicolumn{2}{c|}{MFET} \\  
 \hline
      & $D_1$ & $D_2$ & $T$ ($\delta T$) & $\bar{\T}$ & $\bar{\T}^{*}$ \\  \hline
(i)   &      0 &   1 &     75 (3.1) &   69 &   69  \\
(ii)  &    0.1 & 0.9 &    133 (2.3) &   90 &   90  \\
(iii) &    0.5 & 0.5 &    127 (3.3) &  103 &  103  \\
(iv)  &    0.9 & 0.1 &    134 (4.0) &  107 &  107  \\
(v)   &      1 &   0 &    110 (3.5) &   94 &   94  \\  \hline
\end{tabular}
\caption{
The decay time $T$ and two estimates $\bar{\T}$ and $\bar{\T}^{*}$ of
the MFET from Eqs. (\ref{eq:Tcut}, \ref{eq:Tcut*}), for two diffusing
particles inside a disk of radius $R=10$.  The particle with diffusion
coefficient $D_1$ is initially located at $(0,0)$, whilst the particle
with diffusion coefficient $D_2$ is initially at $(5,0)$.  For cases
(i) and (v), the small-target asymptotic formula (\ref{eq:Tsmall_2d})
yields $T \approx 76.2$ and $T \approx 115.7$, in excellent agreement
with simulation results.  The exact value of $T$ for case (i) is
$72.0$ whereas $\langle \T \rangle$ is $66.6$.  Note that the
estimated value of $T$ here is twice smaller than that from the case
(i) in Table~\ref{Tab:Target_MV} because the diffusion coefficient
$D_1$ was twice smaller in that case.  For case (v), the numerical
solution of the corresponding boundary value problem leads to $T\simeq
108.8$ and $\langle \T \rangle\simeq 92.2$.  }
\label{Tab:DifD_MV}
\end{table}

\subsection*{The limit of very slow targets}
\label{sec:commentMobImmob}

\begin{figure}
\centering
\includegraphics[width=80mm]{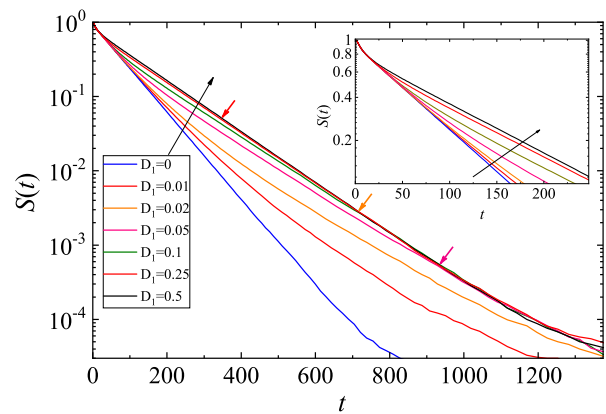} 
\caption{
Survival probability $S(t)$ vs time $t$ for different values of $D_1$:
$0$, $0.01$, $0.02$, $0.05$, $0.1$, $0.25$ and $0.5$, with $D_2 =
1-D_1$.  The first particle starts from the center and the second
particle from $(5,0)$ with $R=10$.  The inset displays the short-time
behavior.  Arrows indicate the time $t_J$ discussed in the text.}
\label{fig:D1overD2to0}
\end{figure}

It is instructive to examine in detail the slow-target limit
$D_1/D_2\to 0$ when the sum of diffusion coefficients is fixed.
Figure~\ref{fig:D1overD2to0} illustrates the behavior of the survival
probability for several values of $D_1$: $0$, $0.01$, $0.02$, $0.05$,
$0.1$, $0.25$ and $0.5$, with $D_2 = 1-D_1$.  For short times
($t\lesssim t_B$), all the lines of $S(t)$ go along the static target
line ($D_1=0$) and, after a certain time, begin to separate from it.
The smaller $D_1$ is, the larger this time becomes.  In this
intermediate time range, $S(t)$ cannot be described as $\propto \exp
(-t/T)$, since it includes a slowly varying prefactor.  After a while,
the lines separate from the target line and bend towards the line $D_1
= D_2 = 0.5$, which is reached (within the resolution of the figure or
simulation errors, that is, within a given relative error) after a
certain time $t_J$.  What we see is that the smaller $D_1$ is, the
larger $t_J$ gets.  These moments are marked with short colored
arrows.  After these times $t_J$, the lines run together, so that
their slope is the same, which means that $T$ is the same, as it
should be.

Since the diffusion operator $D_1 \Delta_{\x_1} + D_2
\Delta_{\x_2}$ depends on both $D_1$ and $D_2$, its eigenvalues and
thus the decay time $T(D_1,D_2)$ is {\it a priori} a function of both
diffusion coefficients $D_1$ and $D_2$.  Even if their sum is fixed,
the decay time is still expected to depend on the ratio $D_1/D_2$.
However, our numerical results and above arguments suggest that, even
for not too large confining volumes, $T(D_1,D_2)$ is, to a very good
approximation, a function of $D_1+D_2$ alone (if $D_1 > 0$ and $D_2
>0$); in this sense, the behavior is reminiscent of the no boundary
case.  This is an important and counter-intuitive result, which
differs from the one-dimensional setting \cite{PartI}, in which the
decay time was indeed a function of both $D_1$ and $D_2$.  In turn,
the characteristic time $t_J$ for relaxation into the monoexponential
regime depends on $D_1$.  In the limit $D_1 \to 0$, $t_J$ seems to
diverge, indicating the singular character of this limit.  In other
words, as $D_1\to 0$, $T(D_1,D_2)$ does not necessarily converge to
$T(0,D_2)$ for the static target.  The singular character of this
limit was established in \cite{PartI} for one-dimensional diffusion on
an interval.  In higher dimensions, it would appear that the above
arguments still carry over, and we therefore conjecture that the
singular behavior would also hold.  However, a more rigorous analysis
is required.  Note, for instance, that the estimates of the decay time
in Table~\ref{Tab:DifD_MV} for the cases with finite diffusivities
differ, although by $5\%$ only.  Concomitantly, we also observe a very
weak dependence of the coefficient $A(D_1/D_2)$ on the ratio $D_1/D_2$
(cf.  Fig.~\ref{Fig:T_VS_Size2D}).

\section{Results in three dimensions}
\label{Sec:3d}

Finally, we briefly extend our study to the three-dimensional case
when two spherical particles of equal radii $\rho_1 = \rho_2 = \rho/2
= 1/2$ diffuse with diffusion coefficients $D_1$ and $D_2$ inside a
sphere of radius $R$ with reflecting boundary.  The results are
qualitatively similar to the two-dimensional case.

The similarity between two- and three-dimensional systems is one of
the most relevant consequences of the presence of reflecting boundary.
In fact, in the no boundary case, two and three dimensional problems
were drastically different.  Even though the MFET is infinite in both
cases, the recurrent Brownian motion performs a compact exploration of
space and visits any infinitesimal region with unit probability,
whereas the transient diffusion in three dimensions may escape to
infinity and never return, in which case encounter never happens.  In
contrast, the boundness of the domain with reflecting boundary makes
diffusion recurrent in any dimension, while the MFET is always finite.
This justifies the similar qualitative behavior in 2D and 3D.  In the
remainder of the section, we undertake a systematic analysis of the
survival probability and of the FET probability density in 3D and
compare them with their 2D counterparts.

\begin{figure}[!t]
\includegraphics[width=0.45\textwidth]{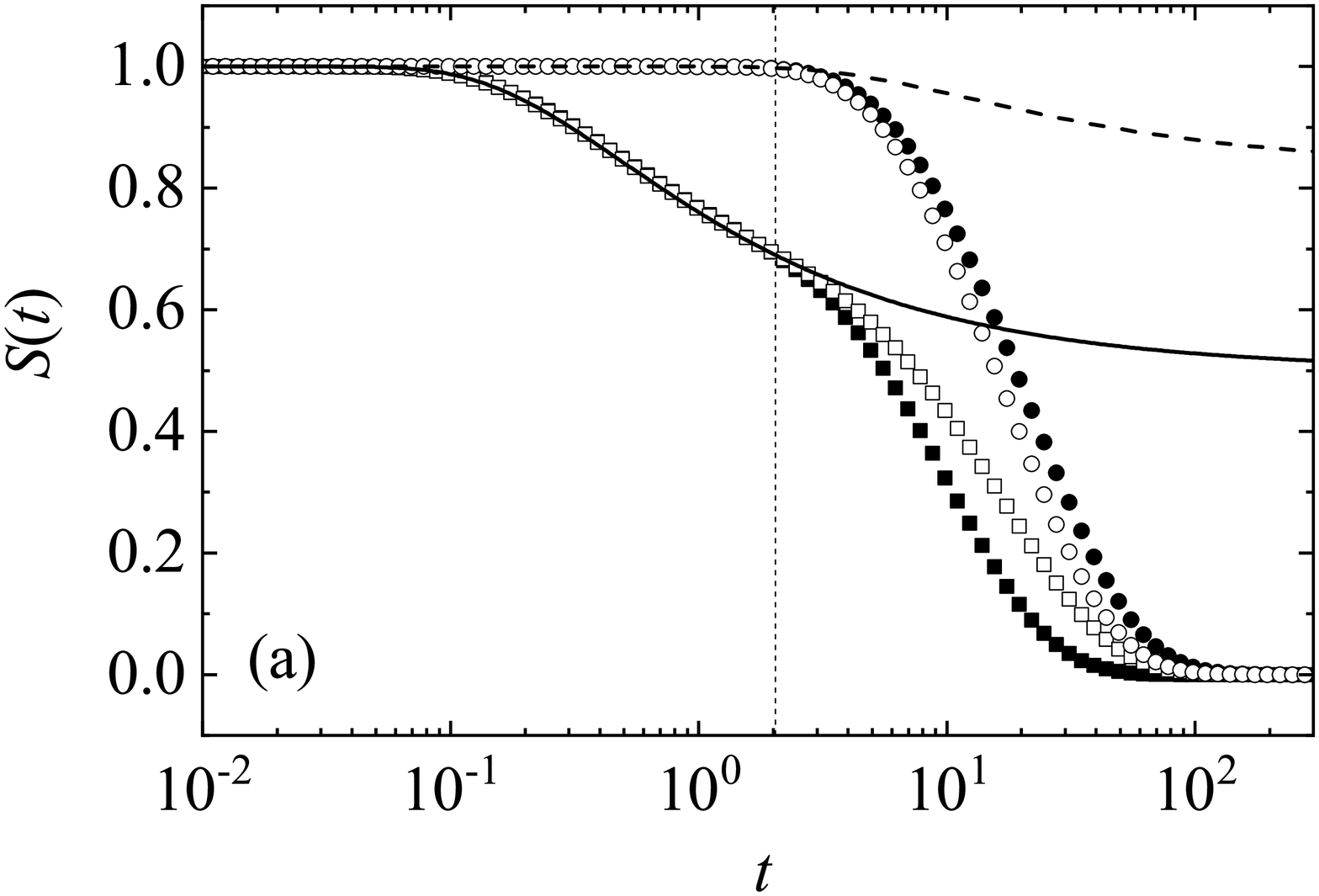} 
\includegraphics[width=0.45\textwidth]{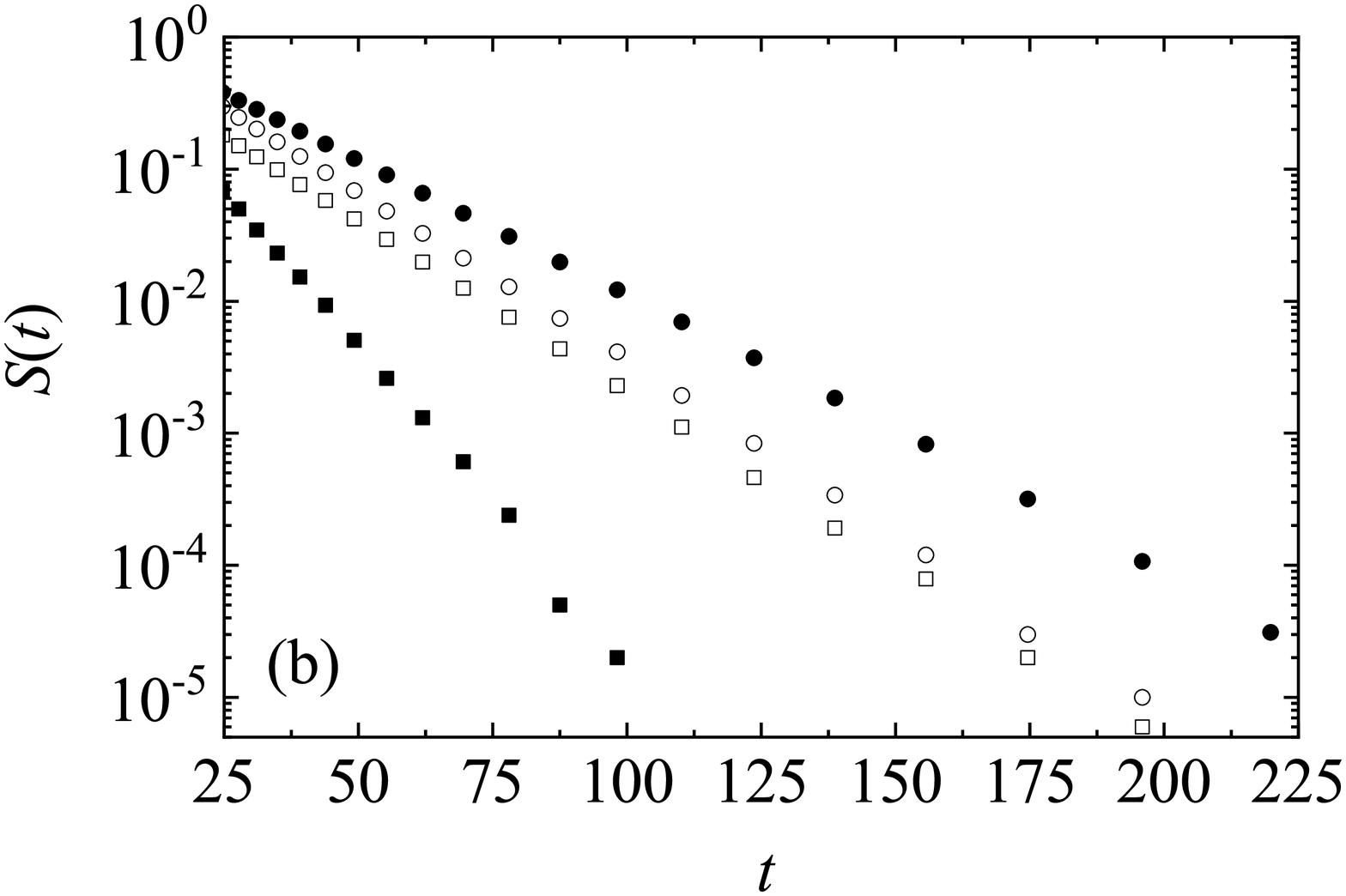} 
\includegraphics[width=0.45\textwidth]{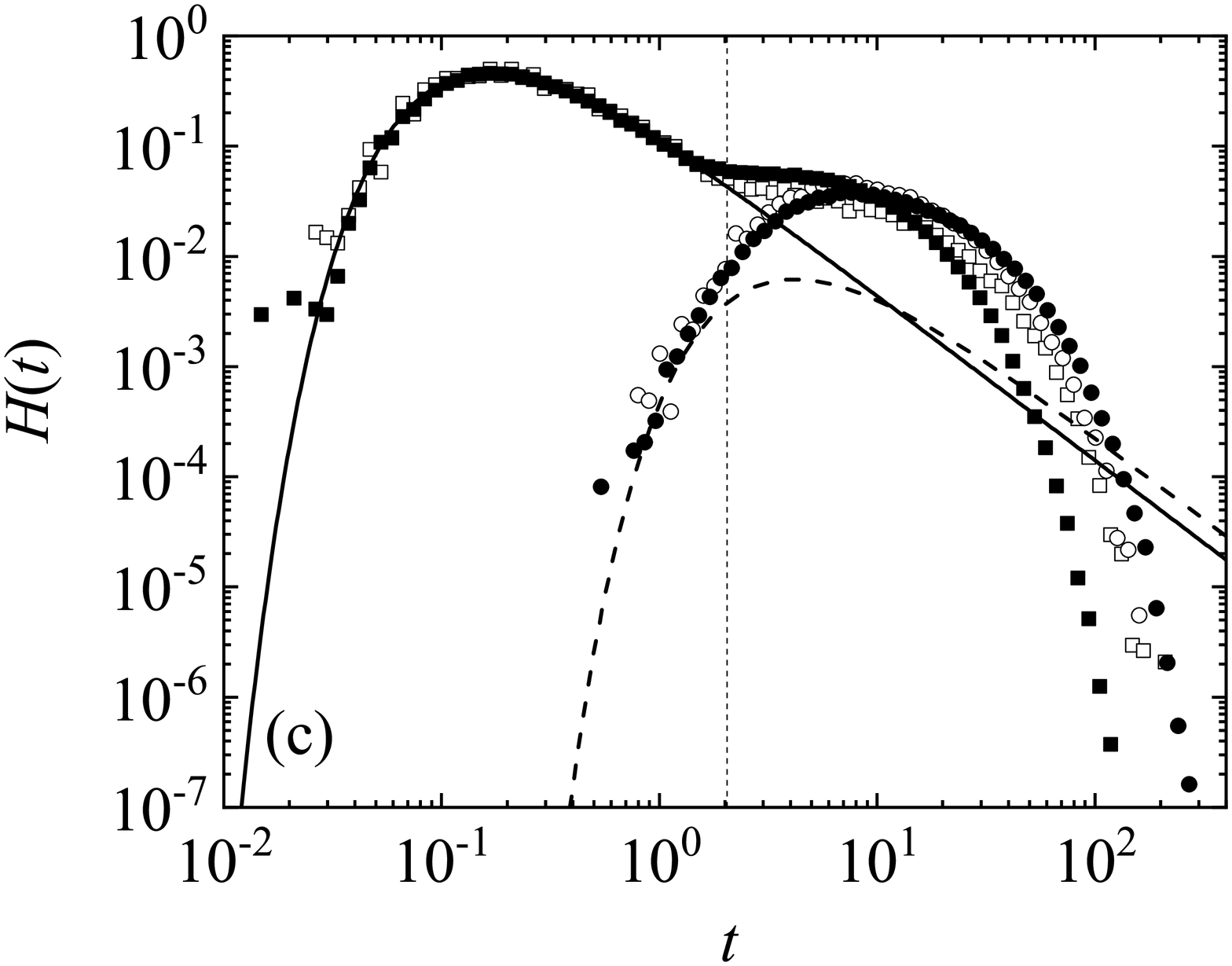} 
\caption{
{\bf (a,b)} Survival probability for two diffusing particles initially
located at $(-r/2,0,0)$ and $(r/2,0,0)$ inside a ball of radius $R =
4$.  Four considered cases are: (i) $r=6$, $D_1=1$ and $D_2=0$ (filled
circles), (ii) $r=6$, $D_1=1/2$ and $D_2=1/2$ (empty circles), (iii)
$r=2$, $D_1=1/2$ and $D_2=1/2$ (empty squares), and (iv) $r=2$,
$D_1=1$ and $D_2=0$ (filled squares).  Lines show $S_{\rm
free}(t|\x_1,\x_2)$ given by Eq. (\ref{eq:S_free3D}) for $r = 6$
(dashed) and $r = 2$ (solid).  Vertical dashed line indicates the
value of $t_B = 2.0$.  Panel {\bf (a)} shows a linear-log plot, whilst
the log-linear representation for long times is presented on panel
{\bf (b)}.  {\bf (c)} FET probability density. }
\label{Fig:SP_3D}
\end{figure}

As in the 2D case, we introduce two timescales $t_B$ and $t_F$ via
Eqs. (\ref{eq:tB}, \ref{eq:tF}).  The survival probability remains
very close to $S_{\rm free}(t|\x_1,\x_2)$ up to $t_B$.  This is
confirmed by Fig.~\ref{Fig:SP_3D}(a), which shows the survival
probability for four different settings (see caption).  At times
around $t_B$, the survival probability for the case (iv) of a fixed
target is lower than that for the case (iii) of diffusing particles,
if the particles are located near the center of the sphere.  In turn,
if both diametrically opposed particles are far from the center, the
case (ii) of diffusing particles favors rapid encounters as compared
to the case (i) of a fixed target.  The explanation is the same as in
2D, where a centered position of a fixed target helps to avoid large
first-encounter times.

Figure~\ref{Fig:SP_3D}(b) illustrates the exponential decay of the
survival probability at long times.  One can see that empty symbols
corresponding to diffusing particles follow two close parallel
straight lines, confirming that the decay time $T$ is independent of
the initial positions.  In turn, filled symbols corresponding to a
fixed target follow straight lines with distinct slopes, highlighting
the dependence of $T$ on the target position.

Figure~\ref{Fig:SP_3D}(c) shows the corresponding FET probability
densities.  Like in the 2D case, the timescales $t_F$ and $t_B$
determine their shapes.  Here, $t_B \simeq 2.0$ for all cases.  Cases
(i) and (ii) present two humps, since $t_F \simeq 0.7 \ll t_B$.  In
contrast, for cases (iii) and (iv) $t_F \simeq 6 \gg t_B$ and the
probability densities present a single hump.  As $T$ is independent on
the initial positions of the particles in cases (ii) and (iii), the
long-time decay of the probability density is roughly the same; in
turn, it is different for cases (i) and (iv), for which $T$ depends on
the initial position of the target (see
Table~\ref{Tab:Identical_MV_3d}).

\begin{figure}[t]
\includegraphics[width=0.45\textwidth]{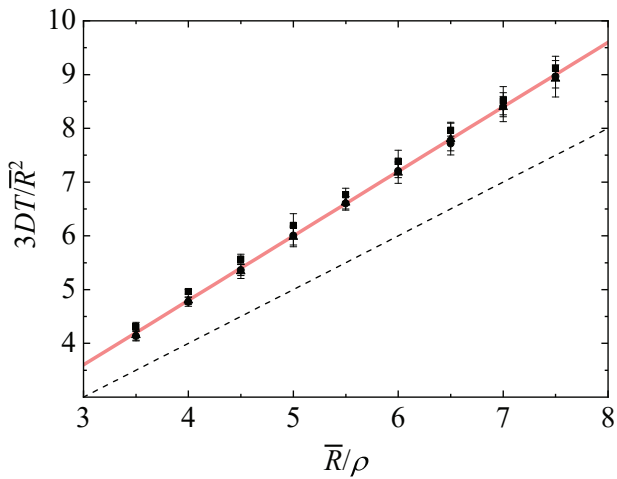} 
\caption{
Scaled decay time $T$ versus the scaled effective radius
$\bar{R}/\rho$ for two diffusing particles started from positions
$(-1,0,0)$ and $(1,0,0)$ inside a confining sphere of radius $R$ equal
to $4,~4.5,~5,~5.5,~6~,~6.5,~7,~7.5$, and $8$, when $D_1=0.9$ and
$D_2=0.1$ (squares), $D_1=D_2=1/2$ (circles), and $D_1=1$ and
$D_2=1/2$ (triangles).  Solid and dashed lines correspond to
Eq.~\eqref{eq:T_3d} with $C_3=1.2$ and $C_3=1$, respectively.  }
\label{Fig:TvsR_3D}
\end{figure}

One important difference with respect to the two-dimensional case is
the dependence of $T$ on the size of the system.  As in 2D, one might
be led to think that Eq. (\ref{eq:Tsmall}), which was obtained for a
fixed small target, is still valid for diffusing particles upon
setting $D=D_1+D_2$.  Figure~\ref{Fig:TvsR_3D} shows the decay time
$T$ as a function of $\bar{R}^3$ for different values of $\bar{R}$ for
two sets of diffusion constants.  In both cases, the scaling of $T$
with $\bar{R}^3$ is confirmed,
\begin{equation}   \label{eq:T_3d}
T \simeq C_3 \frac{\bar{R}^3}{3 D \rho} \,,
\end{equation}
but, in contrast to the prefactor $C_2 \simeq 1$ for the
two-dimensional case, here one has $C_3 \simeq 1.2$.  This result
reflects again the singular character of the limit $D_1/D_2 \to 0$.

\begin{table}
\centering
\begin{tabular}{ |c|c|c|c| }
 \hline
  & Decay time & \multicolumn{2}{c|}{MFET} \\ 
 \hline
 Case & $T$ ($\delta T$) & $\bar{\T}$ & $\bar{\T}^{*}$ \\
 \hline
(i)   &  21.2 (0.3) &  25.4 &  25.4 \\
(ii)  &  16.9 (0.5) &  20.8 &  20.8 \\
(iii) &  16.9 (0.6) &  13.2 &  13.2 \\
(iv)  &   9.5 (0.4) &   8.5 &   8.5 \\  \hline
\end{tabular}
\caption{
The decay time $T$, estimated error $\delta T$, and two estimates
$\bar{\T}$ and $\bar{\T}^{*}$ of the MFET from Eqs. (\ref{eq:Tcut},
\ref{eq:Tcut*}) for two spherical particles, initially placed at
$(-r/2,0,0)$ and $(r/2,0,0)$ inside a ball of radius $R=4$, with (i)
$D_1=1$, $D_2=0$ and $r=6$, (ii) $D_1=D_2=1/2$ and $r=6$, (iii)
$D_1=D_2=1/2$ and $r=2$, and (iv) $D_1=1$, $D_2=0$ and $r=2$.  For
comparison, the small-target asymptotic formula (\ref{eq:Tsmall_3d})
yields $T \approx 9.9$ for the case (ii); in turn, this formula is not
applicable for the case (iv) as the target is too close to the
boundary.  Note that $\rho/\bar{R} \approx 0.29$ is not small, which
can explain discrepancies.  At the same time, Eq. (\ref{eq:T_3d}) with
$C_3=1.2$ yields $T \approx 17.2$ for the cases (i) and (iii), which
differs from the simulation results by less than 2\%.}
\label{Tab:Identical_MV_3d}
\end{table}

\section{Conclusions}
\label{Sec:Conclusions}

In this paper, we studied the distribution of the first-encounter time
for two particles diffusing in bounded domains with reflecting
boundary.  Even though this is a typical situation for many
biochemical reactions, most former studies focused on the much simpler
case with a fixed target ($D_2 = 0$).  This problem of searching for a
fixed target by a single diffusing particle was therefore a reference
benchmark in our analysis, in spite of the singular character of the
$D_2/D_1 \to 0$ limit.  Another benchmark is the no boundary case, for
which the survival probability $S_{\rm free}(t|\x_1,\x_2)$ and the FET
probability density $H_{\rm free}(t|\x_1,\x_2)$ are known explicitly.
The inclusion of a reflecting boundary significantly affects the
survival probability and the FET distribution.  In particular, the
translational invariance of the no boundary case no longer holds.  For
instance, the reflecting boundary makes the survival probability and
the FET distribution explicitly dependent on the initial positions of
the particles with respect to the boundary, not only on the initial
distance between the particles.  This dependence is particularly
significant at short times.  Deviations from $S_{\rm
free}(t|\x_1,\x_2)$ are stronger when both particles are closer to the
boundary.

We introduced two timescales, $t_F$ and $t_B$, that qualitatively
control the FET distribution.  In particular, the survival probability
and the FET probability density can be well approximated by $S_{\rm
free}(t|\x_1,\x_2)$ and $H_{\rm free}(t|\x_1,\x_2)$ when $t \lesssim
t_B$.  In contrast, the confinement effect cannot be generally ignored
at times exceeding $t_B$ In turn, the value of $t_F$ determining the
position of the maximum of $H_{\rm free}(t|\x_1,\x_2)$, affects the
shape of the FET probability density.  When $t_F > t_B$, the FET
probability density exhibits a single maximum and has a mildly broad
shape.  In turn, if $t_F \ll t_B$, the FET probability density is much
broader, and a second hump can emerge at times of the order of $t_B$.

The third important timescale is the decay time $T$ characterizing the
long-time exponential decay of both $S(t|\x_1,x_2)$ and
$H(t|\x_1,\x_2)$.  If both particles are diffusing, the decay time $T$
does not depend on the initial positions of the particles; in turn, if
one particle is immobile (e.g., $D_2 =0$), $T$ depends on its fixed
position.  When the particles are small as compared to the
confinement, our results suggest
\begin{equation}  \label{eq:Tfinal}
T \simeq \frac{C_d \bar{R}^2}{Dd} \times \left\{  \begin{array}{l l} 
 \ln (\bar{R}/\rho) & (d = 2), \\   \bar{R}/\rho & (d=3), \\  \end{array} \right. 
\end{equation}
where $D = D_1 + D_2$, $\bar{R} = R - \rho_1$ and $\rho = \rho_1 +
\rho_2 = 2\rho_1$.  The numerical prefactor $C_2$ was shown to be
close to $1$ in two dimensions, and $C_3 \approx 1.2$ in three
dimensions.  Explaining the deviation of $C_3$ from $1$ remains an
open problem.  In the small-target limit, $T$ is also close to the
MFET.  

Equation (\ref{eq:Tfinal}) can be related to the volume $v(T)$ of the
Wiener sausage generated during time $T$ by a diffusing particle with
diffusion coefficient $D$ and radius $\rho$ \cite{Berezhkovskii1989}.
As long as $D T/\rho^2 \gg 1$, it turns out that Eq. (\ref{eq:Tfinal})
with $C_d=1$ provides the time $T$ required by the diffusive particle
to generate the volume of the Wiener sausage (i.e., the explored
volume up to time $T$) equal to the volume $v(T)$ of the confining
region of radius $\bar{R}$.  This observation allows one to conjecture
how the extension of (\ref{eq:Tfinal}) to $d>3$ could look like: $T
\simeq C_d \bar{R}^d/[d(d-2) D\rho^{d-2}]$, where we used the relation
$v(T)=v_0 d(d-1)DT/\rho^2$ for $d\ge 3$ and $DT/\rho^2$ large, with
$v_0$ being the volume of a $d$-dimensional sphere of unit radius
\cite{Berezhkovskii1989}.

The small-$\rho$ behavior of $T$ described by Eq. (\ref{eq:Tfinal}) is
drastically different from that of the one-dimensional case.  In the
latter, there is no distinction between point-like and finite-size
particles, i.e., the decay time is finite even at $\rho = 0$.  Here,
there is no small-target asymptotic relation like
Eq. (\ref{eq:Tfinal}), and the dependence of the decay time $T$ on the
diffusion coefficients $D_1$ and $D_2$ is not reduced to that of $D_1
+ D_2$ \cite{PartI}.

This work can be extended in several ways.  First, for the sake of
providing realistic descriptions, it is important to investigate the
statistics of first-encounter times in biochemical reactions with
reactants of different sizes.  Even though the same simulation
algorithms can be used, distinct radii add an extra length scale and
thus make the introduction of timescales more subtle.  Second, the
effect of external forces that bias the random motion of diffusing
particles can be important for biological and chemical applications.
Third, one can consider particles undergoing subdiffusive dynamics,
e.g., continuous-time random walks with heavy tailed waiting times
\cite{Montroll1965,Metzler2000}.  As the statistics of particle
trajectories remains unchanged, the subordination concept suggests
that exponential functions in the spectral decomposition of the
survival probability will be replaced by Mittag-Leffler functions,
allowing one to generalize our former results \cite{Grebenkov10b}.
Similarly, one can consider diffusing diffusivity and switching
diffusion models for the dynamics of both particles
\cite{Chechkin17,Lanoiselee18,Grebenkov19f}.  A rigorous mathematical
analysis of the first-encounter distribution in the small target limit
can further clarify the important role of confinement in
diffusion-influenced reactions.  In particular, the derivation of the
leading-order asymptotic relation (\ref{eq:Tfinal}) and the analysis
of its dependence on the diffusion coefficients and the radii of the
particles are still open.

Finally, one of the most important perspectives consists in accounting
for partial reactivity of the particles.  In fact, upon an encounter,
the particles typically have to overcome an energy activation barrier
or to undertake an appropriate conformational change in order to
react.  As a consequence, the reaction occurs with some probability
which depends on the reactivity of the particles.  The role of partial
reactivity in the statistics of first-reaction times of a single
particle diffusing towards a static target was thoroughly investigated
\cite{Collins49,Sano79,Sapoval94,Grebenkov10a,Grebenkov10b,Grebenkov17,Grebenkov19,Grebenkov20a}.
In particular, the concept of the boundary local time characterizing
the number of encounters between the diffusing particle and the static
target was put forward to describe the statistics of the
first-reaction times
\cite{Grebenkov19g,Grebenkov20a,Grebenkov20b,Grebenkov20c,Grebenkov21a,Grebenkov22}.
An extension of the current study to partially reactive particles and
the associated statistics of encounters is of primary importance for a
reliable description of bimolecular reactions.

\begin{acknowledgments}
F. L. V. acknowledges financial support by Junta de Extremadura
(Spain) through Grants GR18079 and PD16010 (partially financed by FSE
funds).  S. B. Y. and E. A. acknowledge financial support from Grant
PID2020-112936GB-I00 funded by MCIN/AEI/10.13039/501100011033, and
from Grants IB20079, GR18079 and GR21014 funded by Junta de
Extremadura (Spain) and by ERDF: A way of making Europe.
D.~S.~G. acknowledges a partial financial support from the Alexander
von Humboldt Foundation through a Bessel Research Award.
\end{acknowledgments}

\appendix
\section{Appendix: Simulation procedure}

In our algorithm, diffusing particles are modeled as continuous-time
random walkers \cite{Montroll1965}.  These particles move randomly by
means of instantaneous jumps.  The motion of each particle in a
$d$-dimensional domain is determined by $d+1$ random variables:
waiting time of a particle until its next jump and its displacements
along each of the $d$ space directions.  These random variables are
drawn from corresponding waiting time and jump length distributions.
In the algorithm, we fix the unit of time by setting the waiting time
PDF to be the exponential distribution $\exp(-t)$.  The random
displacements carried out by the $i$th particle are drawn from
zero-mean Gaussian distribution with variance $\sigma_{ij}^2$, where
$i=1,2$ for two particles, and $j=1, \ldots, d$.  Therefore, the
diffusion coefficient of each particle is equal to
\begin{equation}
D_i = \frac{\sum_{j=1}^d \sigma_{ij}^2}{2d} .
\end{equation}
To deal with isotropic diffusion, we choose $\sigma_{i1}=\ldots =
\sigma_{id}$ in all simulations.  The use of the exponential and
Gaussian PDFs is just a choice; other choices are possible but the
waiting time density should have a finite mean, as well as the jump
length PDF should have a finite variance to produce normal diffusion
\cite{Metzler2000}.

The structure of the program is the following.  At the initial time,
the centers of two particles are set in their prescribed initial
positions.  Then, the times at which the particles are expected to
jump are assigned by means of an exponential random variable.  The
time in the simulation evolves until the minimum of both times.  The
particle with the smaller waiting time takes a jump, whereas the other
particle remains at rest.  The moving particle follows a straight line
from its initial position to its destination.

\begin{figure}[t]
\includegraphics[width=0.45\textwidth]{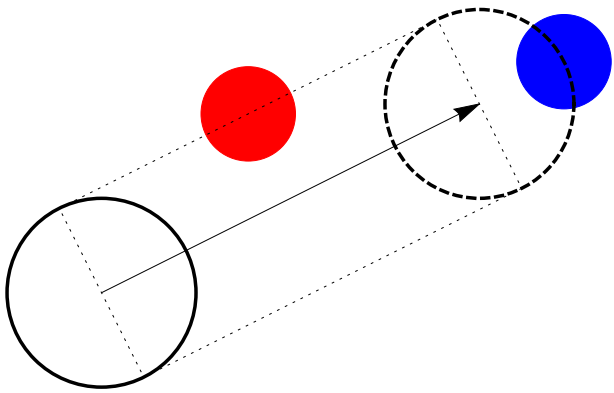} 
\caption{
Illustration of two possible encounters for a $2d$ system in the
simulation algorithm.  The empty disk shows the initial position of
the moving particle, while the dashed circle would be its next
location if there was no encounter.  The arrow indicates the expected
displacement of the center of the moving particle, whereas the
rectangle delimited by dotted segments represents the area swept by
the moving disk during its jump.  The colored disks represent two
possible positions of the static particle that will produce an
encounter event. }
\label{Fig:Reaction_Condition}
\end{figure}

In the no boundary case, only two simple situations could be
distinguished.  If the moving particle collides with the other
particle, the simulation stops and the encounter time is recorded.
Otherwise, the particle arrives at its destination and a new waiting
time is assigned.  The collision takes place if at least one of the
following conditions is fulfilled: (i) the distance from the center of
the static particle (that is, the particle momentarily at rest) to the
destination is smaller than $\rho_1+\rho_2$, where $\rho_1$ and
$\rho_2$ denote respectively the radii of the moving and static
particles; or (ii) there exists a region around the static particle
inside the hypercylinder confined between the initial and final
positions of the moving particle.  Both situations are illustrated in
Fig.~\ref{Fig:Reaction_Condition} for two-dimensional systems (in this
case, the aforementioned hypercylinder is just a rectangle).

However, in the case of bounded domains, the destination may be
outside the confining domain.  It is also possible that the
destination is inside the domain, but its distance to the boundary is
shorter than $\rho_1$.  For our purposes, both situations are
equivalent, since the interaction of any particle of radius $\rho_1$
with a boundary of radius $R$ is the same as that of a point-like
particle with an effective boundary of radius $R-\rho_1$.

The implementation of the reflecting boundary can be done as follows.
Let us assume that the moving particle travels a distance $\Delta l$
in a single step, and that the center of the particle crosses the
effective boundary of radius $R-\rho_1$ after traveling a distance
$\delta l$.  Let $B$ be the point on the line of motion whose distance
to the intersection is equal to $\Delta l - \delta l$, and assume that
this point lies inside the effective disk.  Loosely speaking, let us
also term the radial direction as the line that joins the intersection
with the center of the disk of radius $R-\rho_1$.  Thus, the center of
the moving particle after the jump is the point that is symmetric to
$B$ with respect to the radial direction.  In this case, the encounter
takes place provided that the moving particle collides with the static
particle in the incoming trajectory, or after the reflection.  Also,
there is an encounter if the distance between the final position of
the center of the moving particle and the center of the static
particle is shorter than $\rho=\rho_1+\rho_2$.  Multiple reflections
should be considered when the point $B$ is outside the effective disk.

We also set a time cut-off in order to avoid very long trajectories
prior to the encounter.  The cut-off time is fixed at $t_{\rm
cut}=2500$ for 2D systems and at $t_{\rm cut}=4000$ for 3D.  In all
cases, the number of realizations is $N = 10^6$.

\section{Estimation of the decay time}
\label{sec:estimation}

\begin{figure}
\centering
\includegraphics[width=42mm]{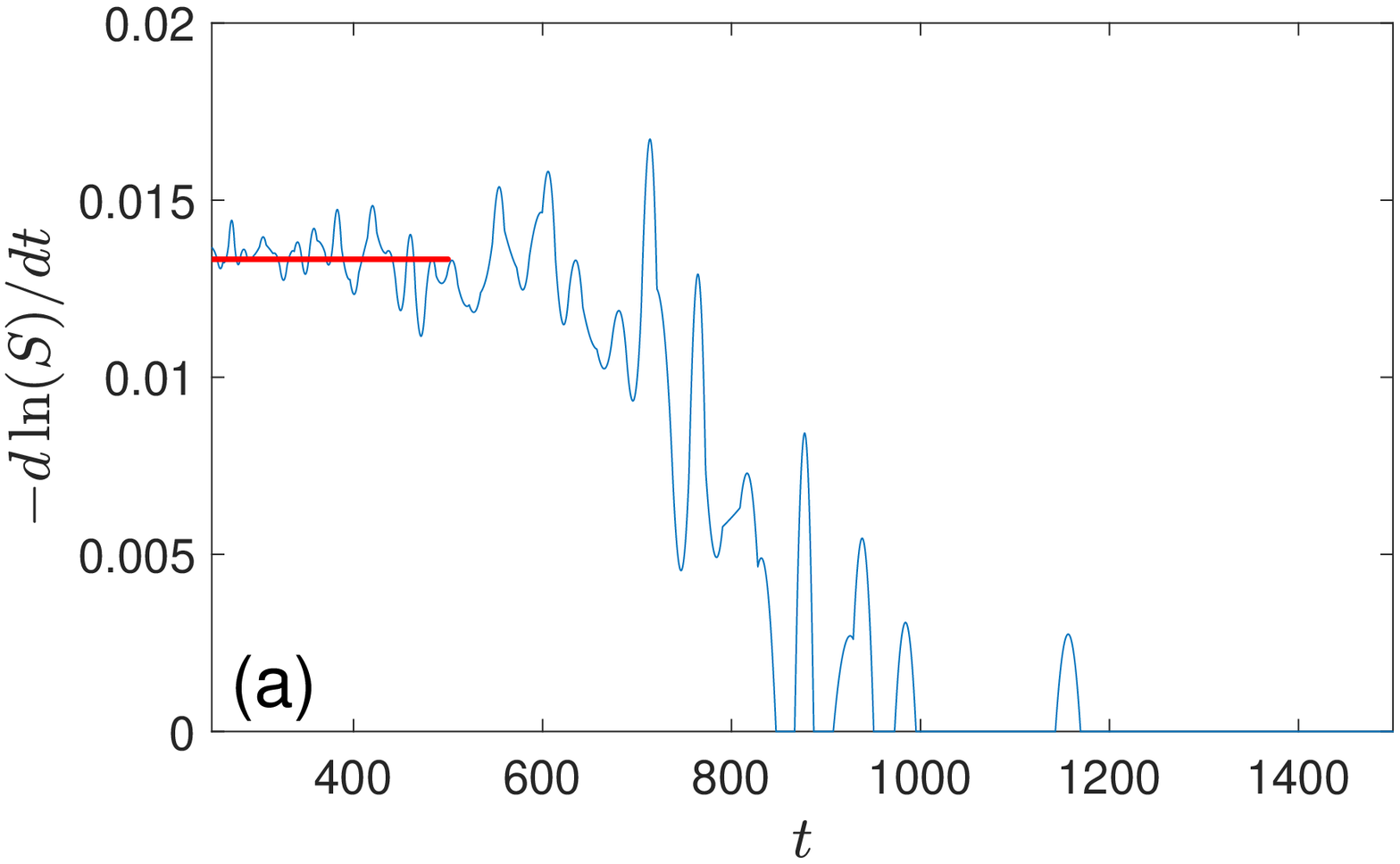} 
\includegraphics[width=42mm]{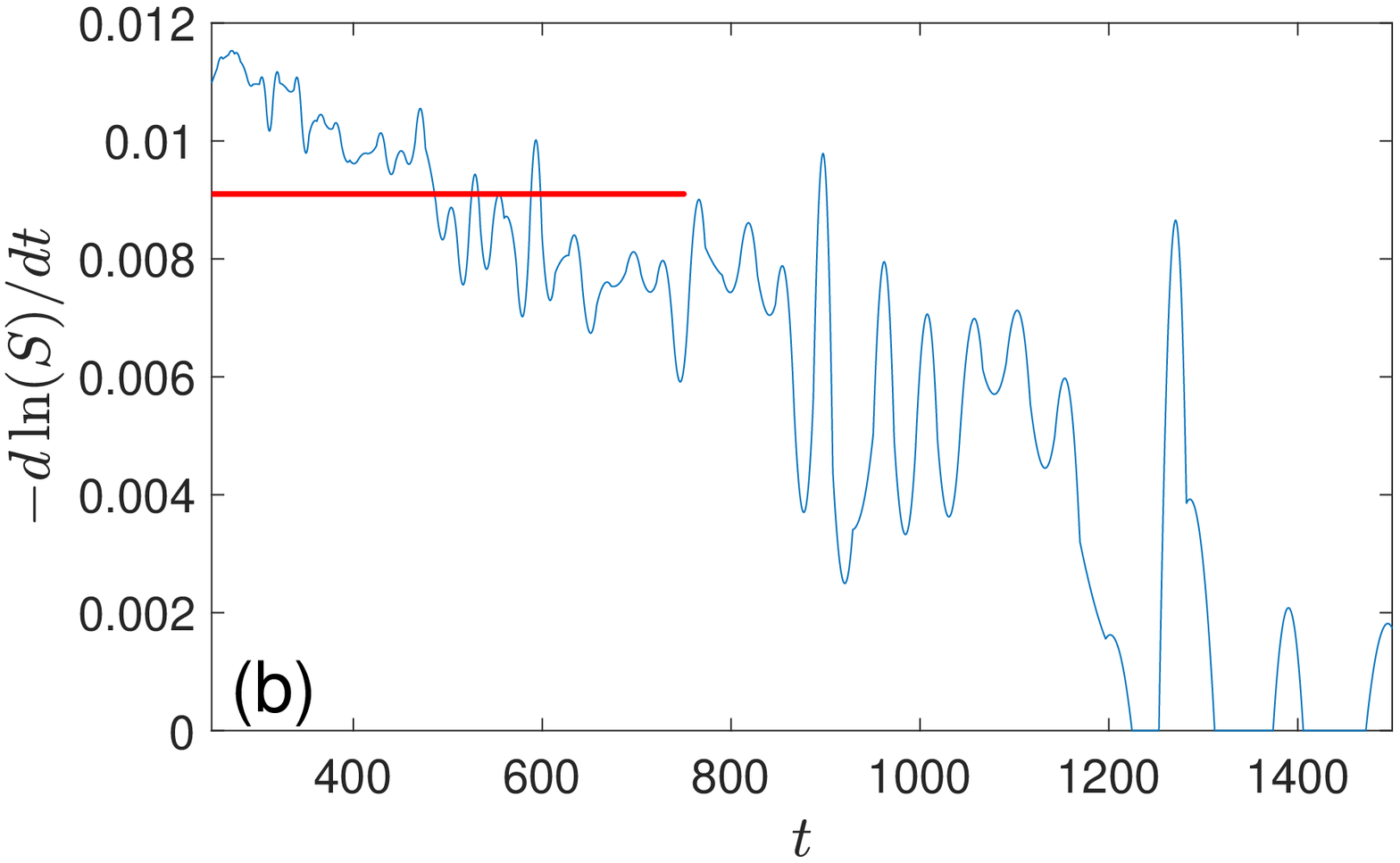} 
\includegraphics[width=42mm]{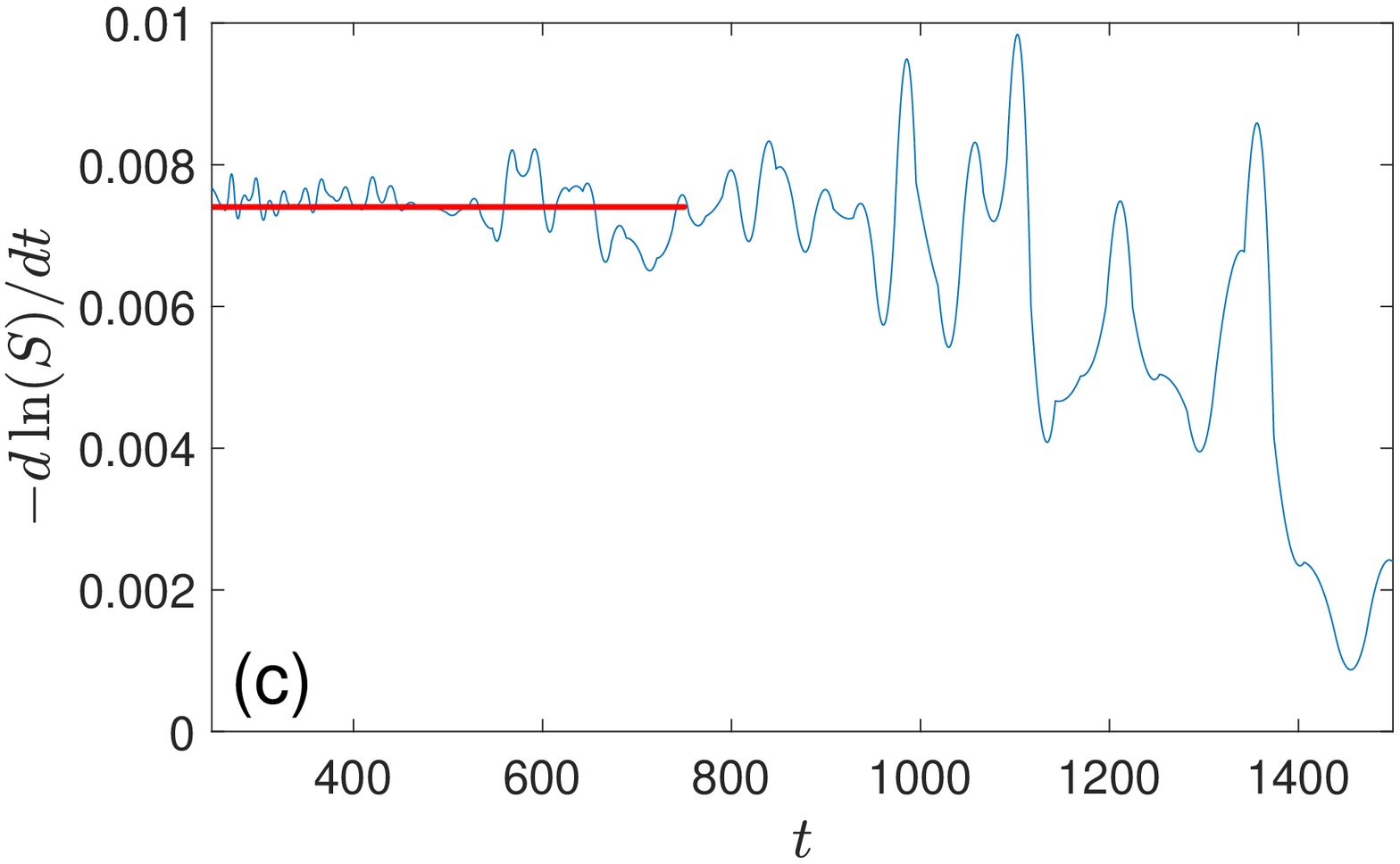} 
\includegraphics[width=42mm]{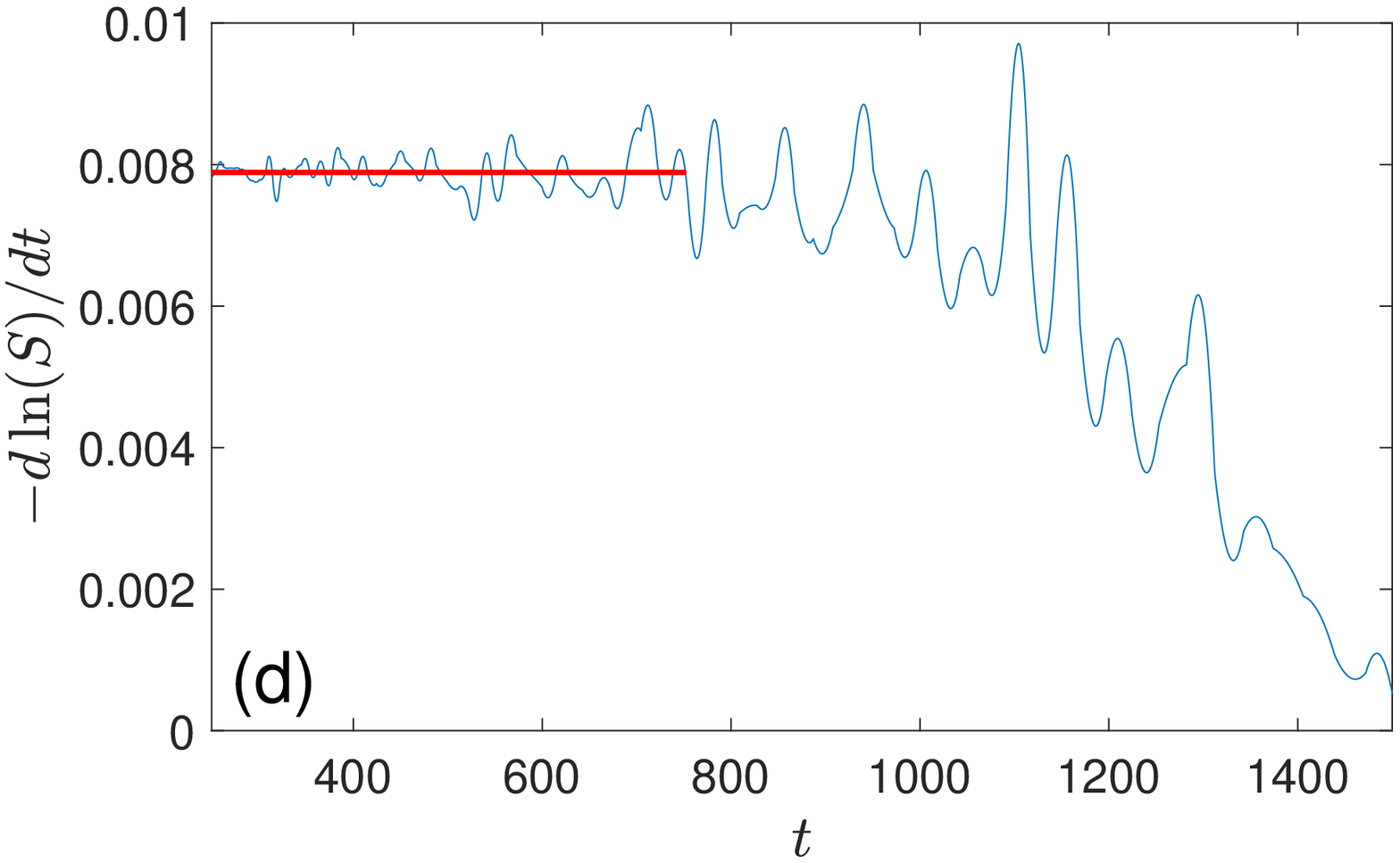} 
\caption{
The logarithmic derivative, $- d\ln S(t)/dt$, of the survival
probability $S(t)$ for 4 choices of $D_1$: $0$ {\bf (a)}, $0.01$ {\bf
(b)}, $0.1$ {\bf (c)}, and $0.5$ {\bf (d)}, with $D_2 = 1-D_1$.  The
first particle starts from the center of the disk of radius $R = 10$
and the second is located at $(5,0)$. Horizontal red line indicates
the range of times $(t_1,t_2)$ used for estimating the decay time $T$.
This estimation fails on the panel {\bf (b)} because the
monoexponential decay arises at longer times, at which the accuracy of
simulations is too low.}
\label{fig:slope}
%
\end{figure}

Estimating the decay time from the long-time asymptotic behavior of
the survival probability is not simple.  As discussed in the main
text, one should carefully select the range of times, $(t_1,t_2)$,
over which the estimation is performed.  In fact, $t$ should be long
enough for the monoexponential decay to already have settled, and
short enough to avoid statistical uncertainties and biases due to a
limited number of Monte Carlo realizations.  Figure \ref{fig:slope}
illustrates this point by showing the logarithmic derivative for 4
choices of $D_1$ (with $D_2 = 1-D_1$), with the slower particle being
at the center of the disk.  For $D_1 = 0$ (fixed target), one observes
a plateau for $t$ from $250$ to $500$, and then a rapid decrease due
to saturation artifacts.  Using this range, one gets the estimate $T
\approx 75$ given in Table~\ref{Tab:DifD_MV}.  Similarly, one gets
accurate estimates of the decay time for $D_1 = 0.1$ and $0.5$.  In
contrast, the logarithmic derivative for the case $D_1 = 0.01$ does
not exhibit a plateau, i.e., the exponential function $e^{-t/T}$ is
affected by another slowly varying function on the considered range of
times.  One therefore needs a larger number of realizations or more
efficient simulation methods (such as in \cite{Nayak20}) to access the
behavior of the survival probability at longer times, for which the
monoexponential decay is well established.

\end{document}